\documentclass[prd,preprint,tightenlines,showpacs,preprintnumbers,nofootinbib,eqsecnum]{revtex4}
 \usepackage[dvips,final]{graphicx}
  \usepackage{amssymb}
   \usepackage{amsmath}
    \usepackage{epsfig}
     \usepackage{bm}
      \usepackage{pifont}
\newcommand{\va}[1]{\langle{#1}\rangle}
\newcommand\convo[1]{\mathop{\otimes}\limits_{#1}}

\begin{document}
\thispagestyle{empty}
\date{\today}
\preprint{\hbox{RUB-TPII-02/07}}

\title{Polarized and unpolarized $\mu$-pair meson-induced Drell--Yan
       production and the pion distribution amplitude\\ }
\author{A.~P.~Bakulev}\email{bakulev@theor.jinr.ru}
\affiliation{Bogoliubov Laboratory of Theoretical Physics, JINR,
               141980 Dubna, Russia}
\author{N.~G.~Stefanis}\email{stefanis@tp2.ruhr-uni-bochum.de}
\affiliation{Institut f\"{u}r Theoretische Physik II,
               Ruhr-Universit\"{a}t Bochum,
               D-44780 Bochum, Germany}
\author{O.~V.~Teryaev}\email{teryaev@theor.jinr.ru}
\affiliation{Bogoliubov Laboratory of Theoretical Physics, JINR,
               141980 Dubna, Russia\\}
\vspace {10mm}

\begin{abstract}
We present a detailed analysis of meson-induced massive lepton
(muon) Drell--Yan production for the process
$\pi^{-}N\to\mu^{+}\mu^{-}X$, considering both an unpolarized nucleon
target and longitudinally polarized protons.
Using a QCD framework, we focus on the angular distribution of
$\mu^+$, which is sensitive to the shape of the pion distribution
amplitude, the goal being to test corresponding results against
available experimental data.
Predictions are made, employing various pion distribution
amplitudes, for the azimuthal angle dependence of the $\mu^{+}$
distribution in the polarized case, relevant for the planned
COMPASS experiment.
QCD evolution is given particular attention in both considered
cases.
\end{abstract}
\maketitle


\section{Introduction}
\label{sec:intro}

The meson-induced production of massive dileptons off baryons in the
Drell--Yan (DY) process \cite{DY70} $MB\to l^{+}l^{-}X$ provides a
useful means of analyzing the quark structure of an unstable hadron,
like the pion.
Indeed, one can extract (see, for example, \cite{Pal85,Con89}) in such 
a context the quark structure function of the pion and test the process
independence of the nucleon structure function measured in deeply
inelastic scattering.

On the other hand, for large $Q^2$ and large longitudinal momentum 
(carried by a pion's quark constituent)
the hadronic differential cross section
for the production of a massive lepton pair via the annihilation of
an antiquark and a quark in the colliding hadrons---pion and nucleon,
respectively---involves the pion distribution amplitude (DA) 
in order to describe 
the pion bound state.
Tuning the DY reaction to the kinematic edge of the phase space,
where the antiquark $\bar{u}$ from the pion is far off-shell, i.e.,
$x_{\bar{u}}\to 1$, it is sufficient to treat the quark $u$,
originating from the nucleon, as being nearly free and on-mass-shell,
so that the bound-state details of the nucleon become irrelevant
\cite{BB79}.
In those circumstances the process $MB\to l^{+}l^{-}X$ reduces to
$\pi^{-}u\to \mu^{+}\mu^{-}X$ and the corresponding amplitude becomes
calculable within perturbative QCD.
The binding effects of the pion state are taken
into account by means of the pion DA (which is the pion wave function
integrated over transverse momenta), making this type of process
suitable to test the details of proposed nonperturbative models for
the pion DA.

Moreover, the angular distribution of the produced lepton pair
(actually the $\mu^+$), relative to the pion direction, depends in a
sensitive way on the pion DA.
Hence, measuring the angular distribution parameters $\lambda,\mu,\nu$
(for an unpolarized target) 
and $\bar\mu, \bar\nu$ 
(for longitudinally polarized protons), 
one can extract useful information on the shape of the pion DA  
\cite{BBKM94,BMT95}.
This is particularly important with regard to a proposed experiment 
by the COMPASS collaboration to collect high-precision data from the 
scattering of a pion beam off a polarized target~\cite{BDP04}.

In the present work we will consider 
(i) the inclusive production of dimuons from the hard scattering of 
    pions on an unpolarized nuclear target
and 
(ii) an analogous process with longitudinally polarized protons.
The appearance of a single transverse-spin asymmetry in the latter case
is related to an imaginary part, which, may even dominate the dimuon 
angular distribution \cite{BMT95}.
Because this contribution is very sensitive to the pion DA, the 
single-spin asymmetry can be used in conjunction with the experimental 
data in order to select the most preferable pion DA.
Our attention is focused in both cases on the role played by the pion 
bound state in terms of the pion DA.

The paper is organized as follows. 
In the next section, we present the theoretical background of the DY 
model in $\pi N$ collisions, taking into account the valence-quark 
bound state of the pion, which we model with the aid of nonlocal QCD 
sum rules \cite{BMS01} in comparison with the Chernyak--Zhitnitsky 
\cite{CZ84} model and the asymptotic pion DA \cite{ER80a,ER80b,LB80}.
Section \ref{sec:comp-data} contains the presentation of our results
for the angular distribution parameters $\lambda, \mu, \nu$, and 
comparison with the data from the E615 experiment at Fermilab
\cite{Con89}.
In this section, we also make predictions for the angular parameters
$\bar\mu$ and $\bar\nu$, using the aforementioned pion DAs.
Particular attention is devoted to the endpoint behavior of
different pion DAs by considering an azimuthal moment of the pion DA,
discussed before in \cite{BMT95}.
Finally, our conclusions are provided in Sec.\ \ref{sec:concl}.

\section{Drell--Yan process of dimuon production}
\label{sec:DY-theory}
This section describes the theoretical method used to calculate the
angular distribution of the $\mu^{+}$ in the DY process in a
$\pi N$ collision in terms of the parameters $\lambda, \mu, \nu$
(unpolarized target) and $\bar{\mu}, \bar{\nu}$ (longitudinally
polarized protons).

\begin{figure}[ht]
 $$\includegraphics[width=0.55\textwidth]{
   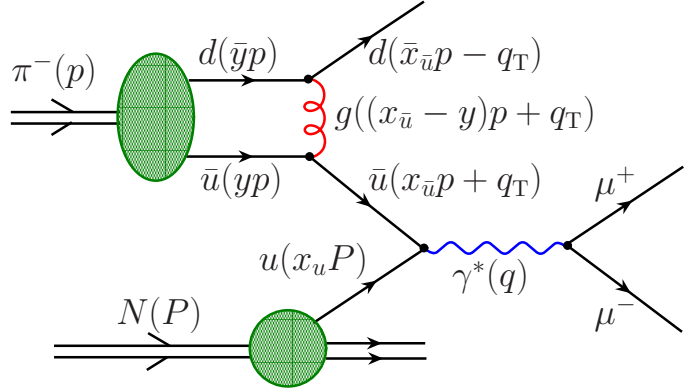}$$
   \vspace{0.0cm} \caption{Graphical representation of the Drell--Yan
   process $\pi^{-}N\to \mu^{+}\mu^{-}X$.
   Symbols $p=p_\pi$, $P=p_N$ and $q=-x_{u}P+x_{\bar{u}}p+q_{T}$,
   where $x_{\bar{u}}=x_{\pi}$, and $x_{u}=x_{N}$ (see for details
   Sec.\ \ref{subsec:kin}), mean four-momenta.
   Here, and below, solid lines denote leptons and also the quarks
   emerging from the $\pi N$ collision, double lines indicate bound
   states (with their corresponding wave functions being illustrated
   by shaded blobs), whereas the (red) curly line stands for the
   exchanged hard gluon and the (blue) weavy line represents the
   highly virtual photon.
   \label{fig:dy-kinem}}
\end{figure}

\begin{figure}[t]
 $$\includegraphics[width=0.45\textwidth]{
    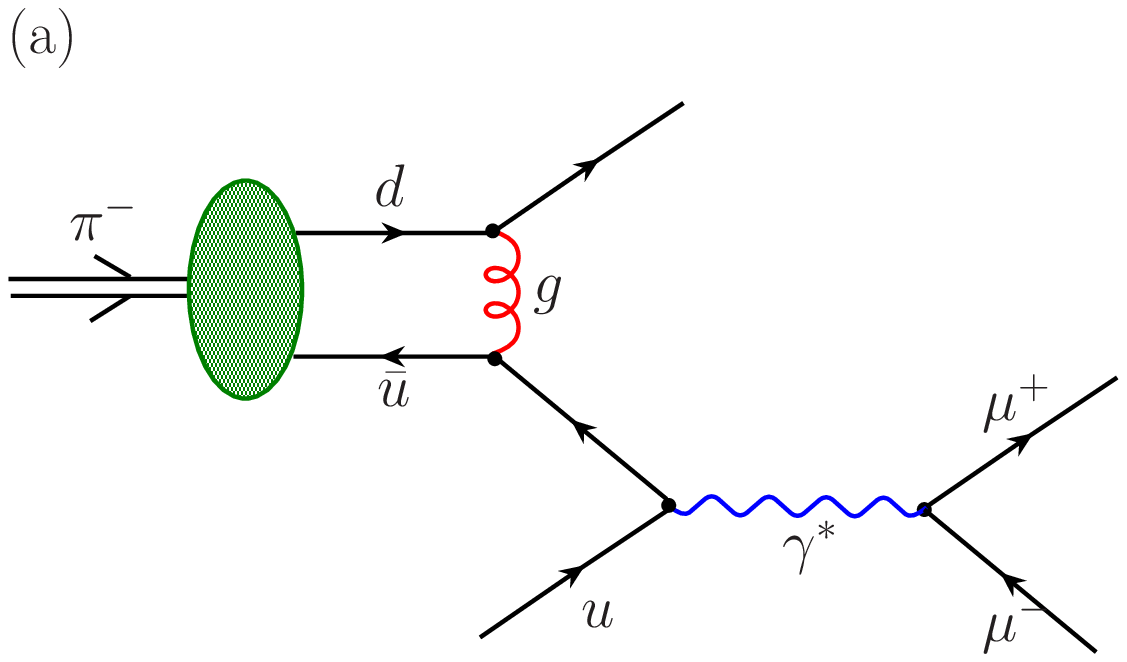}~~~
    \begin{minipage}[c]{0.45\textwidth}\vspace*{-44mm}
     \includegraphics[width=\textwidth]{
     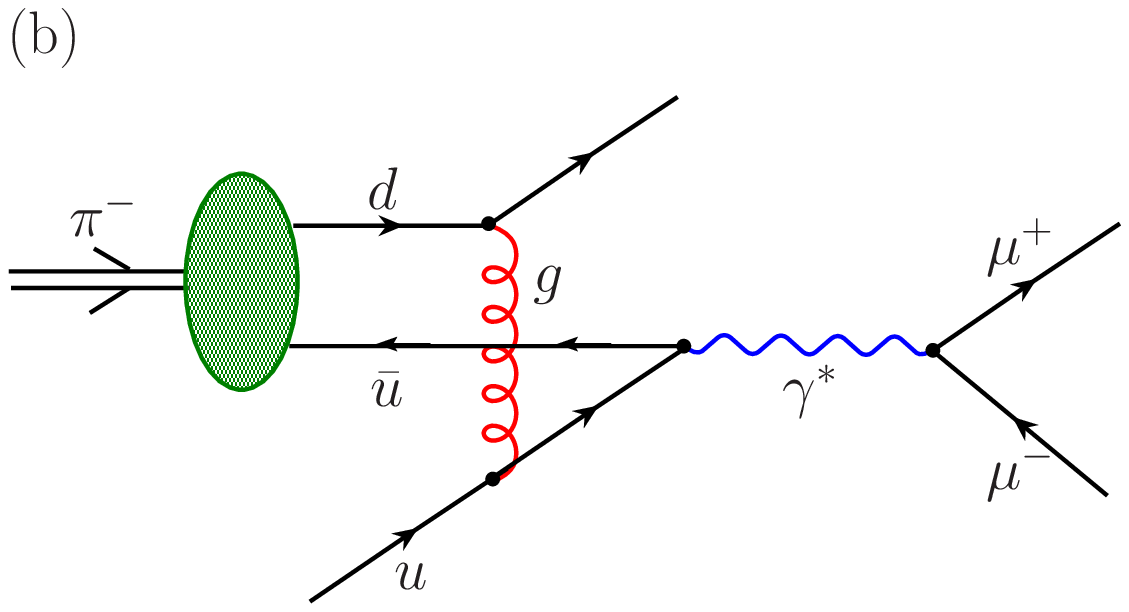}
     \end{minipage}$$
     \vspace{0.0cm} \caption{Leading (perturbative) contributions to
     the QCD-amplitude describing the Drell--Yan process
     $\pi^{-}N\to \mu^{+}\mu^{-}X$ at large $Q^2$ and $x_{L}$, the
     latter defined in Eq.\ (\protect\ref{eq:x_L}).
     The designations are the same as in Fig.\
     \protect\ref{fig:dy-kinem}.
   \label{fig:dy-feyn-diags}}
\end{figure}

The DY process is the dominant mechanism to produce lepton pairs
with a large invariant mass $Q^2$ in hadronic collisions, like
$\pi^{\pm}N$ scattering.
In the context of this model a massive lepton (muon) pair is created
through the electromagnetic annihilation of an antiquark from the beam
pion and a quark from the nucleon target, as depicted in Fig.\
\ref{fig:dy-kinem}.
Moreover, in the kinematic region of large $x_{\pi}=x_{\bar{u}}\to
1$ (or $x_{L}\to 1$, see next section), the antiquark in the pion is
subject to bound-state effects encoded in the valence-quark pion DA,
while the quark from the nucleon is nearly on shell.
Then, the pion bound state can be resolved via hard-gluon exchange
between the annihilating antiquark and the spectator in convolution
with the pion DA \cite{BBKM94}, displayed diagrammatically in
Fig.\ \ref{fig:dy-feyn-diags}.
This provides a means to test the validity of proposed models for the
pion DA, derived from nonperturbative QCD calculations, because the
angular distribution of the produced muon pair depends in a sensitive
way on the shape of the pion DA.
Contributions from higher-order hard-gluon exchange are suppressed by
powers of $\alpha_s$ and will be ignored; those due to evolution will
be taken into account in leading order (LO).
Alternatively, ignoring chromodynamic binding effects, one may extract
in the context of the DY model, the structure functions of the pion and
the nucleon \cite{BB79,Con89} or focus on quark density functions 
\cite{BB79,Ber79}.
These issues are outside the scope of the present investigation.

To continue with our quantitative analysis, we first present some
explanations on the kinematics of the DY process together with the
definitions of the dynamical parameters which describe the angular
distribution in the hadronic differential cross section.

\subsection{Kinematics and variables of the DY process}
\label{subsec:kin}
For the convenience of the reader, the relevant kinematic parameters
and dynamical variables for the DY process $MB\to l^{+}l^{-}X$ (with
$M,B$ denoting a meson and a baryon, respectively) are compiled below
in conceptual groups and in conjunction with
Fig.\ \ref{fig:dy-kinem}:\\

\textbf{Momenta}
\begin{itemize}
\item $p_{\pi}\equiv p\,$: Pion momentum.
\item $p_{N}\equiv P$: Nucleon momentum.
\item $p_{\bar{u}}$: Momentum of the annihilating antiquark $\bar{u}$
      emerging from the pion, $p_{\bar{u}}=x_{\bar{u}}p$.
      As $x_{\bar{u}}\to 1$ (i.e., $\bar{u}$ far off-shell),
      $p_{\bar{u}}^{2}$ becomes large and far spacelike.
\item $x_{\bar u}$: Longitudinal momentum fraction (light-cone
      variable) of the annihilating $\bar u$ from the meson.
      In the kinematical regime, we are considering,
      $x_{\bar u}=x_{\pi}$.
\item $<\mathbf{k}_{T}^{2}>$: Average of the square of the transverse
      momentum of the annihilating $\bar u$ from the meson
      $(\mathbf{k}_{T}^{2}\ll Q^{2})$.
\item $p_{u}$: Momentum of the annihilating quark $u$ emerging from
      the nucleon, $p_{u}=x_{u}P$.
      Because $x_{\bar{u}}\to 1$, it is sufficient to consider the
      $u$-quark to be nearly free and on-shell: $x_{u}=x_{N}$ (quark
      masses and transverse momenta neglected).
\item $Q\equiv m_{\mu\mu}$: Invariant mass of the massive lepton (muon)
      pair, or, equivalently, momentum of the virtual photon
      created by the annihilated quarks $(Q^2=q^{2})$.
      In this analysis we consider values in the range
      $Q^2=(4-81)$~{\rm GeV}$^2$.
\item $s$: Squared invariant mass of the initial hadrons, i.e.,
      $s=(p+P)^2$.
      In our numerical analysis we use $s=100-400$~\text{GeV}$^2$,
      which covers the $252$~GeV $\pi^{-}$ beam of the E615
      Collaboration at Fermilab \cite{Con89}.
\end{itemize}

\textbf{Masses}
\begin{itemize}
\item $m_{\pi}$: Pion mass.
      Because $s\gg m_{\pi}^2$, we set $m_{\pi}=0$.
\item $m$: Bare quark mass.
      Because $s\gg m^2$, we set $m=0$;
\item $m_{N}$: Nucleon mass.
\end{itemize}

\textbf{Kinematic Variables}
\begin{itemize}
\item $\tau\equiv Q^2/s$: Scaling parameter.
\item $\rho\equiv Q_{T}/Q$: This scaling parameter is a measure
      for the squared transverse momentum
      $Q_\text{T}^2\equiv-q_\text{T}^2$ of the virtual photon
      in the hadronic center-of-mass frame (c.m.f.).
      Inspection of Fig.\ \ref{fig:dy-kinem} reveals that
      $q_\text{T}$ appears only in combination with $x_{\bar{u}}p$,
      namely, $q_\text{T}+x_{\bar{u}}p$.
      Then, in order to separate both terms unambiguously, one has
      to demand that $q_\text{T}$ should not contain any part of
      the pion momentum $p$, cf.\ Eq.\ (\ref{eq:q_T.a_T}).
\item $x_{L}=2Q_{L}/\sqrt{s}<1$: Longitudinal momentum fraction
      (Feynman $x$) of the lepton pair (associated with the virtual
      photon) in the hadron c.m.f.\ $(Q_{L}^{2}=q_{L}^2)$. $Q_{L}$ and
      $Q_{T}$ are the photon-momentum components parallel and
      perpendicular, respectively, to the incident pion momentum
      in the hadron c.m.f.
\item $x_{F}=x_{\pi}-x_{N}$ and $x_{\pi}x_{N}=\tau$.
      Neglecting the quark transverse momentum and mass (as $s$ becomes
      very large), the quantities $x_{\pi}$ and $x_{N}$ can be
      identified with $x_{\bar{u}}$ and $x_{u}$, respectively.
      Combining the two equations above, one finds \cite{Con89}
      $x_{\pi,N}=
      \left[\pm x_{F} + \left(x_{F}^{2} + 4\tau\right)^{1/2}\right]/2$.
\end{itemize}

\textbf{Angular distribution parameters}
\begin{itemize}
\item $\theta$: Polar angle measuring the $\mu^{+}$ direction with
      respect to the $t$-channel (or Gottfried--Jackson system of
      axes), i.e., $\cos\theta=\hat{p}_{\mu}\cdot\hat{p}_{\pi}$.
      The definitions of other choices of axes (frames) can be found,
      e.g., in \cite{Con89}.
\item $\phi$: Azimuthal angle between the massive lepton-pair plane and
      the plane of the incident hadrons in the lepton rest frame.
\item $\lambda$, $\mu$, $\nu$: Angle-independent coefficients,
      depending on $Q^2$ and $x_L$.
      These three parameters control the dilepton angular distribution
      and are sensitive to the shape of the pion DA.
\item $\bar\mu$, $\bar\nu$: Angle-independent coefficients, depending
      on $Q^2$ and $x_L$, induced by the imaginary part to the DY
      amplitude in the polarized case.
      They are sensitive to the shape of the pion DA.
\end{itemize}

In our analysis we consider values of $s$ much larger than the nucleon
mass, $s\geq 100\, m_N^2$, so that this be neglected, though we
derive the exact result with $m_N^2\neq0$ and proceed then with the
approximate expression with $m_N=0$, where it is applicable.
The momenta $p$ and $P$ are on the light-cone and, hence, we have
\begin{eqnarray}
 p^2\ =\ 0\,,\qquad
 P^2\ =\ m_N^2\ \approx 0\,,\qquad
 2(p\cdot P) &=& s - m_N^2\ \equiv\ \tilde{s}\
             \approx\ s \,.
\end{eqnarray}
For the cross-section calculation, 
one can appeal to the optical theorem and set the upper right line 
in Fig.\ \ref{fig:dy-kinem}, denoting the $d$-quark, 
on mass shell.
Then,
\begin{eqnarray}
 \label{eq:2pq_T}
  2(p\cdot q_\text{T})
   &=& \frac{q_\text{T}^2}{1-x_{\bar{u}}}
    =  \frac{-\rho^2 Q^2}{1-x_{\bar{u}}}
\end{eqnarray}
and, due to $q=(x_{u}+x_{\bar{u}})p+q_\text{T}$, 
one obtains
\begin{eqnarray}
  Q^2 &=& \left(s-m_N^2\right)\,x_{u}\,x_{\bar{u}}\,
          \left[1+\frac{x_{u}\,m_N^2}{x_{\bar{u}}\,(s-m_N^2)}\right]
        - \frac{\rho^2\,Q^2}{1-x_{\bar{u}}}\,
           \left(1+\frac{2\,x_{u}\,m_N^2}{s}\right)\nonumber\\
      &\approx&
          s\,x_{u}\,x_{\bar{u}}\,
          \left(1+\frac{x_{u}\,m_N^2}{x_{\bar{u}}\,s}\right)
        - \frac{\rho^2\,Q^2}{1-x_{\bar{u}}}\,,~~~
\end{eqnarray}
or
\begin{eqnarray}
  x_{u}\,x_{\bar{u}}
   &=& \left[1 + \frac{\rho^2}{1-x_{\bar{u}}}
   \left(1+\frac{2\,x_{u}\,m_N^2}{s}\right)\right]\,
        \frac{Q^2}{\tilde{s}+m_N^2(x_{u}/x_{\bar{u}})}
 \nonumber\\
   &\approx&
       \left(1 + \frac{\rho^2}{1-x_{\bar{u}}}\right)\,
        \tau\,
         \left(1 - \frac{x_{u}\,m_N^2}{x_{\bar{u}}\,s}\right),~~~
\end{eqnarray}
which fixes the value of $x_{u}$ as a function of
$x_{\bar{u}}$, $\rho$ and $\tau$.
We also see that a non-zero nucleon mass is important only
at very small $x_{\bar{u}}\simeq m_N^2/s$.
At this value of $x_{\bar{u}}$ the value of $x_u$ is close to 1,
so that, neglecting $m_N$, one finds 
$1/\left[1+(m_N^2/Q^2)/(1+\rho^2)\right]$.
Hence, the difference can be sizable and, especially at
$Q^2\sim2m_N^2$ and $\rho\sim1$, it can reach as much as 20\%.

In the following considerations 
(and, in particular, in the analysis of the polarized DY process), 
we will use the longitudinal-momentum fraction of the photon, $x_L$.
To this end, recall that in the hadron c.m.f. one has
\begin{eqnarray}
  p = \frac{\sqrt{s}}{2}(1,0,0,+1)\,,\ \
  P = \frac{\sqrt{s}}{2}(1,0,0,-1)\,,\ \
  q_{\perp} = (0,q_{\perp1},q_{\perp2},0)~~~
\end{eqnarray}
and, using Eq.\ (\ref{eq:2pq_T}), one finds
\begin{eqnarray}
  q_{T} &=& q_{\perp} - a_{T}\,P\,,\ \
  a_{T}\,\equiv\,\frac{\rho^2 \tau}{1-x_{\bar{u}}}\,,
  \label{eq:q_T.a_T}\\
  q &=& q_{\perp} + q_{L}\,,\ \
  q_{L}\,=\,x_{\bar{u}}\,p + (x_{u}-a_{T})\,P
            \,=\,\frac{\sqrt{s}}{2}(x_0,0,0,x_{L})\, ,
\end{eqnarray}
with $x_{L}$ being defined by
\begin{eqnarray}
  x_0\,=\,x_{\bar{u}} + x_{u} - a_{T}\,,\ \
  x_{L}\,
       =\,x_{\bar{u}} - x_{u} + a_{T}\,.
\end{eqnarray}
It is convenient to recast $x_{L}$ in terms of
$x_{\bar{u}}$, $\rho$, and $\tau$ to read
\begin{eqnarray}
  x_{L}
   &=& x_{\bar{u}} - \frac{1+\rho^2}{x_{\bar{u}}}\,\tau\,.
\label{eq:x_L}
\end{eqnarray}
The inverse relation
\begin{eqnarray}
  x_{\bar{u}}
   &=& \frac{x_{L} +\sqrt{x_{{L}}^2+4(1+\rho^2)\tau}}{2}\
\end{eqnarray}
defines $x_{\bar{u}}$ as a function of $x_L$, $\rho$, and $\tau$.

\subsection{Drell--Yan $\pi N$ process with a pion bound state}
\label{subsec:DY-pion}
There have been attempts \cite{BBKM94} and, in particular \cite{BMT95},
to test the compatibility of different pion DAs with the DY angular
distribution measured in the scattering of pions off protons.
The first work suggested that the CZ pion DA 
(and some variant of it) 
fits the unpolarized data \cite{Con89} better than the asymptotic DA, 
or narrower convex versions \cite{Ber79}.
The second work focused on the possibility of longitudinally 
polarized protons as a target and discussed additional angular 
parameters that vanish for an unpolarized target.
It was argued that future polarized experimental data on these 
processes may be able to distinguish among various pion DAs 
because of the high sensitivity of these parameters to the particular 
shape of the pion DA.
Though there is still no such data available, meanwhile important 
knowledge has been collected that is rather unfavorable for the 
endpoint-dominated type of two-humped pion DAs, like the CZ one.
For instance, new high-precision lattice simulations
\cite{DelD05,Lat05,Lat06} give a new level of detail for the second 
Gegenbauer coefficient $a_2$, yielding values around 0.2 to 0.24 at a 
momentum scale of 2~GeV, well within the range suggested by the 
nonlocal QCD sum-rule estimates \cite{BMS01}---hereafter referred to 
as BMS---and supported by the analysis in \cite{BMS02,BMS03,BMS05lat} 
of the CLEO data \cite{CLEO98} on the $\pi-\gamma$ transition form 
factor using light-cone QCD sum rules.
This $a_2$ value is about two times smaller than its counterpart of
the CZ pion DA and cast serious doubts about the consistency of this
model with experiment.
Moreover, the aforementioned BMS CLEO-data analysis has confirmed
the earlier Schmedding and Yakovlev \cite{SY99} findings which
excluded the CZ pion DA at least at the $2\sigma$ level.
On the other hand, the asymptotic pion DA seems to be also excluded,
given that its Gegenbauer coefficient $a_2$ is identically zero and 
the CLEO-data analysis relegates this DA outside the $2\sigma$ error 
ellipse (for the most recent rigorous analysis, see \cite{BMS05lat}), 
while being also incompatible with the lattice results 
\cite{DelD05,Lat05,Lat06}.
Further theoretical arguments and details can be found in
\cite{BMS04kg}.
See also~\cite{BP06zako} for a recent development of the nonlocal
QCD sum-rule approach and the extraction of the pion DA.

Therefore, it would seem reasonable and timely to upgrade the
calculation of the DY angular parameters of the $\pi^{-}N$
hard-scattering process by taking into account the recent
developments quoted above.
To continue, we first recall the definition of the pion DA,
$\varphi_{\pi}(y,\mu_{0}^2)$, which specifies the fractional
longitudinal momentum $y$ of the valence-quark constituents in the
pion at the normalization scale $\mu_{0}^2$.
At the (leading) twist-two level it is defined by the following 
matrix element
\begin{eqnarray}
 \label{eq:pi-DA-ME}
 \va{0\mid\bar{d}(z)\gamma^{\mu}\gamma_5\,
 {\cal C}(z,0) u(0)\mid\pi(P)}
  \Big|_{z^2=0}
 &=& i P^{\mu}\,f_{\pi}
      \int^1_0 dy\,e^{iy(zP)}\,
      \varphi_{\pi}\left(y,\mu_{0}^{2}\right)\ ,\\
 \int_0^1 \varphi_{\pi}(y,\mu_{0}^{2})\, dy
  &=& 1\,,
 \label{eq:pi-DA-fpi}
\end{eqnarray}
where $f_{\pi} = 130.7 \pm 0.4$~MeV \cite{PDG2002} is the pion decay
constant defined by
\begin{equation}
\langle 0|\bar{d}(0)\gamma_{\mu} \gamma_{5} u(0)|\pi^{+}(P) \rangle
 =
ip_{\mu} f_{\pi}.
\label{eq:fpi}
\end{equation}
Above, a straight path-ordered Fock--Schwinger connector \cite{Ste84} 
(Wilson line)
$
{\cal C}(0,z)
  = {\cal P}
  \exp\!\left[-ig_s\!\!\int_0^z t^{a} A_\mu^{a}(y)dy^\mu\right]
$
has been inserted to preserve gauge invariance of the operator product.
In the following, we use the light-cone gauge which reduces the 
contribution of the Wilson line to unity.
The normalization scale, $\mu_{0}^{2}$, 
of the pion DA is related to the ultraviolet (UV) regularization 
of the quark-field operators on the light cone 
in (\ref{eq:pi-DA-fpi}), whose product becomes singular for $z^2=0$.

The pion DA cannot be derived from first principles, 
but has to be inferred from nonperturbative QCD models 
(or from experiment).
This not withstanding, its evolution is governed by perturbative QCD
\cite{ER80a,ER80b,LB80} and can be expressed in the form
\begin{eqnarray}
 \varphi_{\pi}(y, \mu^{2})
  = U(y,s; \mu^{2},\mu_{0}^{2})\convo{s}\varphi_{\pi}(s,\mu_{0}^{2})\,, 
  \qquad \qquad  
  \convo{s} \equiv \int_0^1 ds\,,
 \label{eq:PhipVP}
\end{eqnarray}
where $\varphi_{\pi}(s, \mu_{0}^{2})$ 
is a nonperturbative input determined 
at some low-energy normalization point $\mu_{0}^{2}\sim 1$~GeV${}^2$, 
where the local operators in Eq.\ (\ref{eq:pi-DA-ME}) are renormalized, 
while $U(y,s; \mu^{2}, \mu_{0}^{2})$ is the evolution operator from 
that scale to the observation scale $\mu$, calculable in QCD 
perturbation theory.
In the asymptotic limit, the shape of the pion DA is completely fixed
by pQCD to be $\varphi_{\pi}^{\rm asy}(y)=6y(1-y)$ \cite{ER80a,ER80b,LB80}, 
with the nonperturbative information being solely contained 
in the pion-decay constant $f_{\pi}$.

\begin{figure}[t]
 \centerline{\includegraphics[width=0.5\textwidth]{
              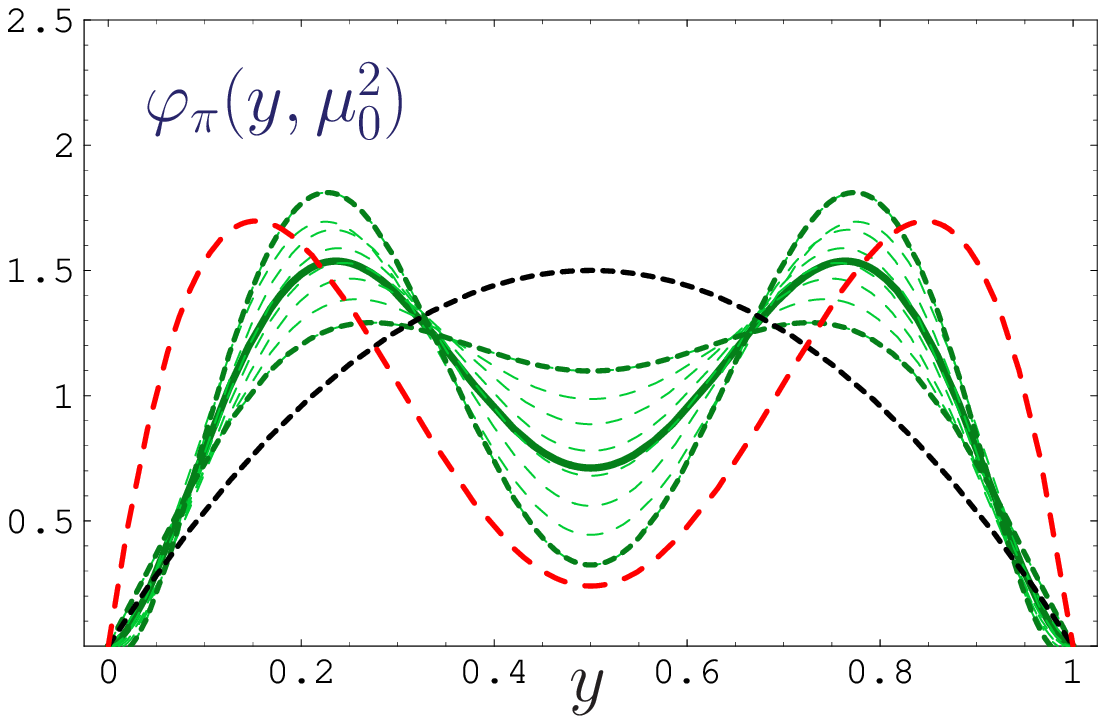}
             \includegraphics[width=0.5\textwidth]{
              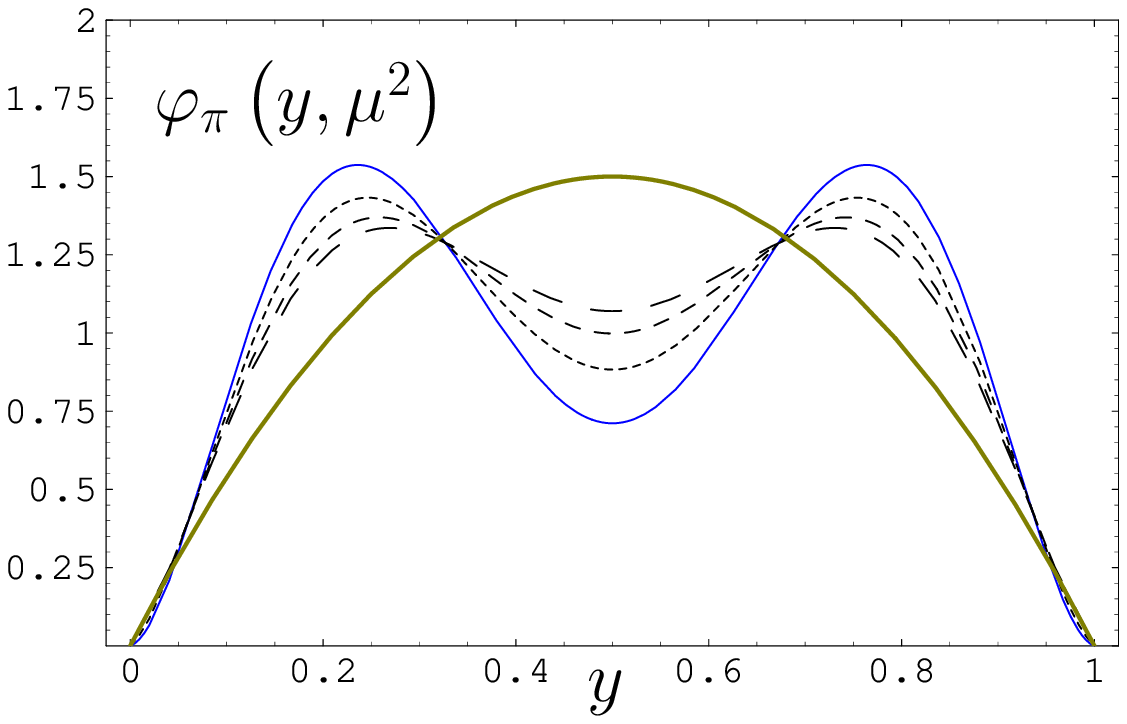}}
  \caption{\footnotesize The left panel shows the ``bunch'' of pion DAs
   (green broken lines), derived from nonlocal QCD sum rules (for a
   summarized exposition, see, e.g., \protect\cite{BMS04kg}), in
   comparison with two extreme alternatives: the asymptotic DA 
   \protect\cite{ER80a,ER80b,LB80}---dotted line---and the CZ model 
   \protect\cite{CZ84}---red long-dashed line---at the momentum scale
   $\mu^2 \approx 1$~GeV$^{2}$.
   The green solid line inside the ``bunch'' represents the BMS model
   \protect\cite{BMS01}.
   The right panel illustrates the effect of one-loop evolution on the
   pion DA, exemplified by the BMS model, in comparison with the
   asymptotic solution (continuous convex line).
   The double-humped solid line represents
   $\varphi_\text{BMS}^\text{LO}(x)$ at 1 GeV$^2$,
   while the  broken lines mark $\varphi_\text{BMS}^\text{LO}(x)$
   at 4, 20, and 100 GeV$^2$ (with the larger scale corresponding
   to the larger value of the DA at the middle point).
\label{fig:pion_das}}
\end{figure}

In the leading-twist approximation of the pion DA, in which we are 
working, $\varphi_\pi(y,\mu_0^2)$ can be expressed in terms of the 
Gegenbauer polynomials, which form an orthonormal set of 
eigenfunctions.
Then,
\begin{eqnarray}
 \varphi_\pi(y,\mu_0^2)
  = 6 y (1-y)
     \left[ 1
          + a_2(\mu_{0}^{2}) \, C_2^{3/2}(2 y -1)
          + a_4(\mu_{0}^{2}) \, C_4^{3/2}(2 y -1)
          + \ldots
     \right]\,,
\label{eq:phi024mu0}
\end{eqnarray}
with all nonperturbative information being encapsulated in the 
expansion coefficients $a_n$.
Depending on the nonperturbative approach applied, these coefficients 
can be calculated via QCD sum rules \cite{CZ84}, 
QCD sum rules with nonlocal condensates \cite{BMS01}, 
or lattice simulations \cite{DelD05,DelDebbio05,Lat05,Lat06}.
More details can be found in the original papers already cited,
while we here restrict ourselves to the obtained results, 
which we quote in Table \ref{tab:Gegenbauers}.
The shapes of these DAs are displayed in the left panel of 
Fig.\ \ref{fig:pion_das}, whereas the evolution effect is illustrated 
in the right panel of this figure in terms of the BMS pion DA.
Note that the underlying quark virtuality employed, is 
$\lambda_q^2=0.4$~GeV${}^{2}$ \cite{BMS02}
corresponding to a correlation length of the scalar quark nonlocal 
condensate of about $0.31$~fm.
As one anticipates from this figure, the key characteristic of the 
BMS-type pion DAs is that the endpoint region $x\to 0,1$ is strongly 
suppressed---not only relative to the CZ pion DA, but even with 
respect to the asymptotic one (for mathematical details, see 
\cite{BMS04kg}).
In contrast, the double-humped shape---reminiscent of the CZ pion
DA---turns out to be of minor importance \cite{BPSS04}.

\begin{table}[h]
\caption{Pion DA models used in the analysis.
\label{tab:Gegenbauers}}
\begin{ruledtabular}
\begin{tabular}{ccccc}
$\pi$--DAs$|_{\mu^{2}=1~{\rm GeV}^2}$~~
        & asymptotic~~
        & BMS \protect\cite{BMS01}~~
        & BMS ``bunch'' \protect{\cite{BMS01,BMS03}}~~
        & CZ \protect\cite{CZ84}\
             \\ \hline \hline
$a_2$  & $0$
       & $0.20$
       & $[0.13, 0.25]$
       & $0.56$
              \\
$a_4$  & 0
       & $-0.14$
       & $[-0.04, -0.22]$
       & 0
             \\
higher
       & 0
       & negligible
       & negligible
       & 0 \\
\end{tabular}
\end{ruledtabular}
\end{table}

Let us now make some remarks on the evolution of the pion DA in LO
of perturbative QCD.
Taking into account only the first two Gegenbauer coefficients,
one obtains
\begin{eqnarray}
 \varphi_{\pi}^{\text{LO}}(y,\mu_\text{F}^2)
  = 6 y (1-y)
     \left[ 1
          + a_2^{\text{LO}}(\mu_\text{F}^2) \, C_2^{3/2}(2 y -1)
          + a_4^{\text{LO}}(\mu_\text{F}^2) \, C_4^{3/2}(2 y -1)
     \right]\,,
\label{eq:phi024LO}
\end{eqnarray}
where $a_2^{\text{LO}}(\mu_\text{F}^2)$ and 
$a_4^{\text{LO}}(\mu_\text{F}^2)$
are given by
\begin{eqnarray}
 a_n^{\text{LO}}(\mu_\text{F}^2)
  &=& a_n(\mu_{0}^{2})\,
      \left[\frac{\alpha_{s}(\mu_\text{F}^2)}{\alpha_{s}(\mu_{0}^{2})}
      \right]^{\gamma_n^{(0)}/(2b_0)}\,.
 \label{eq:anLO}
\end{eqnarray}
The expressions for the anomalous dimensions $\gamma_n^{(0)}$ and the
beta-function coefficient $b_0$ are listed, for example, in 
\cite{BMS02}.
One sees from the right panel of Fig.\ \ref{fig:pion_das} that the
effect of the inclusion of the LO diagonal part of the evolution 
kernel is indeed important.
This figure shows how the BMS pion DA
$\varphi_\text{BMS}^\text{LO}(x)$ 
evolves from the normalization scale of 1 GeV$^2$ (double-humped 
solid line) to higher momentum values at 4, 20, and 100 GeV$^2$ 
(broken lines), with the larger scales corresponding also to larger 
values of the DA at the middle point.
The asymptotic profile (continuous solid line) is displayed for
comparison.

\subsection{DY reaction with an unpolarized target}
\label{subsec:unpol}
The angular distribution of the $\mu^{+}$ in the pair rest frame
can be written in terms of the kinematic variables $\lambda, \mu, \nu$
as follows
\begin{eqnarray}
  \frac{d^{5}\sigma(\pi^{-}+N
  \to\mu^{+}+\mu^{-}+ X)}{dQ^{2}dQ_{T}^{2}dx_{L}\,d\cos\theta d\phi}
&& \propto
  N(\tilde{x},\rho)\Big(1 + \lambda \cos^{2}\theta +
  \mu \sin 2\theta \cos \phi
\nonumber \\
&&   ~~~~~~~~~~~~~~~~\, 
+ \frac{\nu}{2} \sin^{2}\theta \cos 2\phi\Big) \, ,
\label{eq:dif-cross}
\end{eqnarray}
where \cite{BBKM94}
\begin{eqnarray}
 \lambda(\tilde{x},\rho)
  &=& \frac{2}{N}
    \bigg\{\left(1-\tilde{x}\right)^2
           \big[(\textbf{Im}\,I(\tilde{x}))^2
                +(F+\textbf{Re}\,I(\tilde{x}))^2
           \big]
         - (4-\rho^2)\,\rho^2\,\tilde{x}^2\,F^2
    \bigg\}\,,
 \label{eq:lambda.unp}\\
 \mu\left(\tilde{x},\rho\right)
  &=& -\frac{4}{N} \rho\,\tilde{x}\,F\,
     \bigg\{\left(1-\tilde{x}\right)\,
            \big[F+\textbf{Re}\,I(\tilde{x})\big]
            +\rho^2\,\tilde{x}\,F
     \bigg\}\,,
 \label{eq:mu.unp}\\
 \nu\left(\tilde{x},\rho\right)
  &=& -\frac{8}{N}\,\rho^2\,
        \tilde{x}\,\left(1-\tilde{x}\right)\,
         F\,\big[F+\textbf{Re}\,I(\tilde{x})\big]\,,
 \label{eq:nu.unp}\\
 N\left(\tilde{x},\rho\right)
  &=& 2\,\bigg\{\left(1-\tilde{x}\right)^2
              \big[\left(\textbf{Im}\,I(\tilde{x})\right)^2
                  +\left(F+\textbf{Re}\,I(\tilde{x})\right)^2
              \big]
            +(4+\rho^2)\,\rho^2\,\tilde{x}^2\,F^2
         \bigg\}
 \label{eq:norm.unp}
\end{eqnarray}
with
\begin{eqnarray}
 \label{eq:tilde.x}
  \tilde{x}\left(x_L,\rho\right)
   \equiv \frac{x_L + \sqrt{x_L^2+4(1+\rho^2)\tau\vphantom{^|}}}
               {2\left(1+\rho^2\right)}\,.
\end{eqnarray}
Also displayed is the normalization factor of the cross section, $N$.
The abbreviations
\begin{eqnarray}
 F &=& \int_0^1 dy\,
        \frac{\varphi(y,\tilde{Q}^2)}{y}\,
 \label{eq:inv.mom}\\
 I\left(\tilde{x}\right)
   &=& \int_0^1 dy\,
        \frac{\varphi(y,\tilde{Q}^2)}
             {y\,\left(y+\tilde{x}-1+i\varepsilon\right)}
 \label{eq:mom.I}
\end{eqnarray}
are functionals of the pion DA and, therefore, depend on the 
evolution momentum scale $\tilde{Q}^2\sim Q^2$.
Note that the inverse moment (\ref{eq:inv.mom}) of the pion DA plays a
crucial role in the description of several form factors of the pion in
perturbative QCD \cite{BPSS04}.
We will have more to say about the choice of the evolution scale in the
next subsection.
Note also that the denominators in Eqs.\ (\ref{eq:inv.mom}) and
(\ref{eq:mom.I}) originate from the gluon and quark propagators in the
subprocesses shown in Fig.\ \ref{fig:dy-feyn-diags}, respectively.
Before presenting results for these angular coefficients, let us first
discuss what changes are induced when the nuclear target is polarized.

\begin{figure}[t]
 $$\includegraphics[width=0.45\textwidth]{
    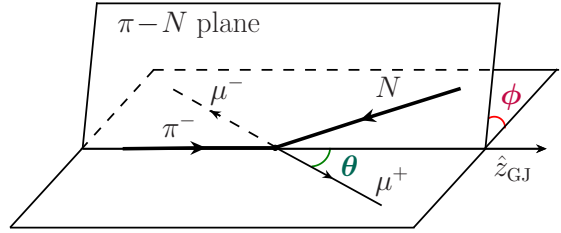}$$
   \vspace{0.0cm} \caption{Angular definitions of the Drell--Yan 
   process in the center of mass frame of the produced massive lepton 
   pair.
   The axis $\hat{z}_{\rm GJ}$ denotes the pion direction in the
   Gottfried--Jackson (GJ) frame.
   \label{fig:dy-angles}}
\end{figure}

\subsection{DY reaction with longitudinally polarized protons}
\label{subsec:pol}
In the polarized DY process the angular distribution of the $\mu^+$
contains two additional parameters $\bar\mu$ and $\bar\nu$,
entering additively Eq.\ (\ref{eq:dif-cross}) with the same angular
structure as $\mu$ and $\nu$, respectively, being, however,
proportional to the target longitudinal polarization $s_{\ell}\,$:
\begin{eqnarray}
 \label{eq:mu.bar}
  \bar{\mu}\left(\tilde{x},\rho\right)
   &=& \frac{-2\,\pi\,s_\ell\,\rho\,\tilde{x}\,
       F\,\varphi(\tilde{x},\tilde{Q}^2)}
            {\left(1-\tilde{x}\right)^2\,
              \left[\left(F+\textbf{Re}\,I(\tilde{x})\right)^2
                  + \pi^2\,\varphi(\tilde{x})^2\right]
                  + (4+\rho^2)\,\rho^2\,\tilde{x}^2\,F^2}\,
       \bar{\mu}_\text{nucl}\,,\\
  \bar{\mu}_\text{nucl}
   &\equiv&
        \frac{{4\over 9}\,\Delta q_u^v(x_p;\mu^2)
            + {4\over 9}\,\Delta q_u^s(x_p;\mu^2)
            + {1\over 9}\,\Delta q_d^s(x_p;\mu^2)}
             {{4\over 9}\,q_u^v(x_p;\mu^2)
            + {4\over 9}\,q_u^s(x_p;\mu^2)
            + {1\over 9}\,q_d^s(x_p;\mu^2)}\,,
 \label{eq:mu.bar.nucl}\\
 \bar{\nu}\left(\tilde{x},\rho\right)
   &=& 2 \rho\,\bar{\mu}\left(\tilde{x},\rho\right)\,.
 \label{eq:nu.bar}
\end{eqnarray}
where $x_p=\tau/\tilde{x}$, $\mu^2$ is the evolution scale for the
nucleon parton distributions.
Note that all momenta refer to the hadronic c.m.f.
The polarized parton distributions used in our analysis are taken
from \cite{GS94}, whereas for the unpolarized structure functions 
we use the parameterization of Ref.\ \cite{GRV95}.
To evolve these distributions from their normalization scale
$\mu_{0}^{2}=4$~GeV${}^{2}$ to the scale $\mu^2=Q^2=16$~GeV${}^{2}$,
we employed the Fortran codes supplied by the authors of these papers
on the Durham web site.\footnote{%
\textsl{http://durpdg.dur.ac.uk/hepdata/grv.html} --- for GRV95
and\\
\textsl{http://durpdg.dur.ac.uk/pdflib/gehrmann/pdf/welcome.html} ---
for GS96.}

Lacking radiative corrections to the process under study, we cannot
fix the evolution scale unambiguously.
In an effort to get a measure for the entailed uncertainty, we have 
analyzed the dependence of our results on the choice of the evolution 
scale by varying $\mu^2$ in the range
$\mu^2=Q^2/2$ and $\mu^2=2 Q^2$.
We found that, depending on the model pion DA used, the variance of
the calculated $\bar{\mu}$ parameter lies between 2\% and 17\%,
with moderate sensitivity to the adopted $\rho$ value.
Details are given in Table \ref{tab:Evol_Error} for the (large) $x_L$
interval, in which the one-gluon exchange is still a good
approximation and only the pion bound-state is of importance.
Recall in this context that the only evolution effect in the case of 
the asymptotic pion DA stems exclusively from the DGLAP evolution 
of the nucleon parton distributions.
On the other hand, in the case of the BMS and the CZ pion DAs, 
the total evolution effect is the result of the combination of the 
ERBL evolution of the pion DA and  the DGLAP evolution of the nucleon 
parton distributions 
(unpolarized and polarized) in $\mu_{\rm nucl}$ 
(cf.\ Eq.\ (\ref{eq:mu.bar.nucl})).

\begin{table}[h]
\caption{Change of predictions for $\bar{\mu}(x_L,\rho)$, shown in
Fig.\ \ref{fig:rho-evo-2}, due to different settings of the evolution
scale.
\label{tab:Evol_Error}}
\begin{ruledtabular}
\begin{tabular}{cccc}
 $\pi$-DAs
            & asymptotic
                   & BMS \protect\cite{BMS01}
                         & CZ \protect\cite{CZ84}\
                              \\ \hline
$\rho=0.06$ and $0.5\leq x_L\leq0.84$
            & $2.0-2.5$\%
                   & $12-2.5$  \%
                         & $17-2.5$ \% \\
$\rho=0.30$ and $0.5\leq x_L\leq0.83$
            & $2.0-2.5$\%
                   & $9-0$  \%
                         & $13-0$ \% \\
$\rho=0.50$ and $0.5\leq x_L\leq0.81$
            & $1.5-2.0$\%
                   & $8-0$  \%
                         & $13-8$ \% \\
\end{tabular}
\end{ruledtabular}
\end{table}

It was noted in \cite{BMT95} that the angular moment of the pion DA,
defined by
\begin{eqnarray}
 \label{eq:Az-Mom-Def}
  {\cal M}_\text{ang}
   = \int \sin 2\theta\, \sin\phi\,
            d\sigma(s_{\ell}=1)
   = -2\,\pi\,\rho\,\tilde{x}\,F\,\varphi(\tilde{x},\tilde{Q}^2)\,
       \bar{\mu}_\text{nucl}\,,
\end{eqnarray}
is particularly sensitive to the $\tilde{x}$ (or $x_L$) endpoint
region and can be used in comparison with experimental data in
order to distinguish pion DAs which behave differently exactly in
this region.
We have, therefore, included predictions also for this quantity 
(see next section, Fig.\ \ref{fig:rho-evo-Ang-Mom}).
In similar context, it is important to consider the (experimental) 
single-spin azimuthal asymmetry (SSA)
\begin{eqnarray}
 \label{eq:SSA}
 {\cal A}
  \equiv \frac{\displaystyle d\sigma(s_{\ell}=+1)-d\sigma(s_{\ell}=-1)}
              {\displaystyle d\sigma(s_{\ell}=+1)+d\sigma(s_{\ell}=-1)}
\end{eqnarray}
after averaging the cross sections over the polar angle
$\theta\in[0,\pi]$:
\begin{eqnarray}
 \label{eq:SSA_Mu_Nu_La}
  {\cal A}\left(\phi,x_L,\rho\right)
   = \frac{\displaystyle \rho\,\bar{\mu}(s_{\ell}=+1)\sin 2\phi}
          {2 + \lambda + \frac12\,\nu\cos 2\phi}\,.
\end{eqnarray}
In Eq.\ (\ref{eq:SSA}), \hbox{$\sigma(s_{\ell}=+1)$} 
and \hbox{$\sigma(s_{\ell}=-1)$}
denote opposite helicity states of the longitudinally polarized
target.
Notice that in order that the longitudinally polarized nucleon 
parton distributions can transfer its polarization to the azimuthal 
distribution of the massive lepton pair, one needs an imaginary part 
and the interference of amplitudes with a phase difference between
them.
Both ingredients are provided here by the pion DA and the hard-gluon 
exchange.
Ignoring pion bound-state effects in the treatment of the DY process, 
one would have to include in the hard cross sections perturbative QCD 
radiative corrections proportional to $\alpha_s$ in order to create 
a SSA.

\begin{figure}[t]
 $$\includegraphics[width=0.32\textwidth]{
    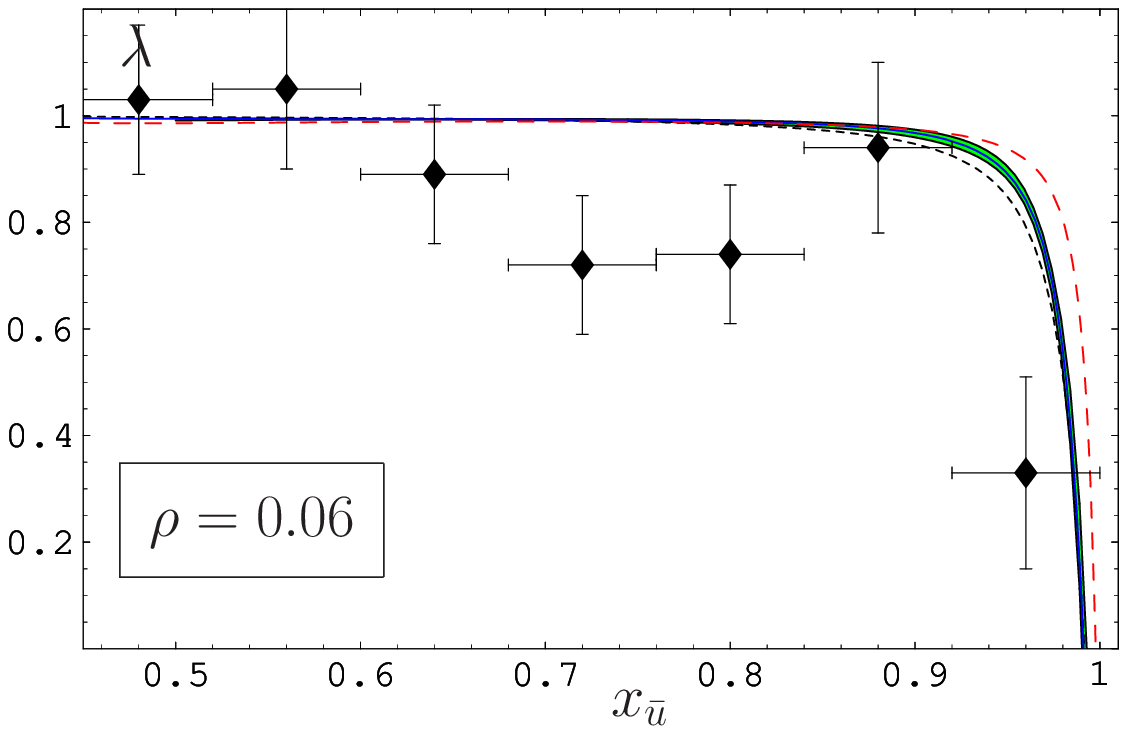}~~%
   \includegraphics[width=0.32\textwidth]{
    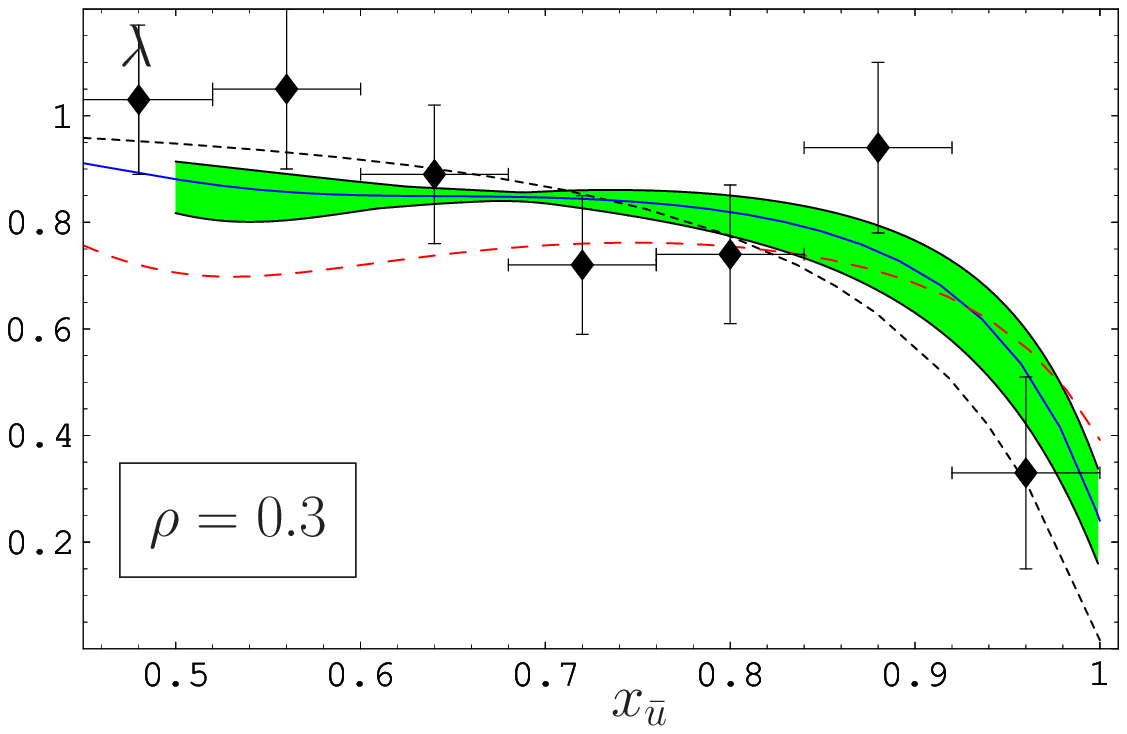}~~%
   \includegraphics[width=0.32\textwidth]{
    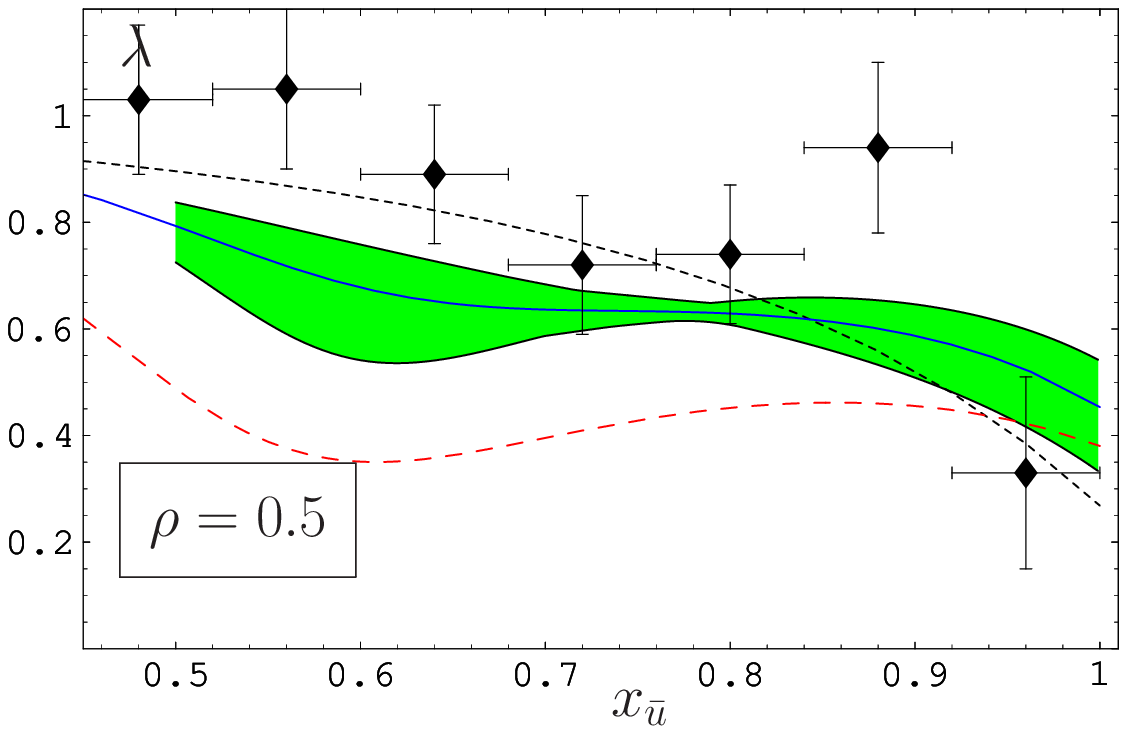}$$
 $$\includegraphics[width=0.32\textwidth]{
    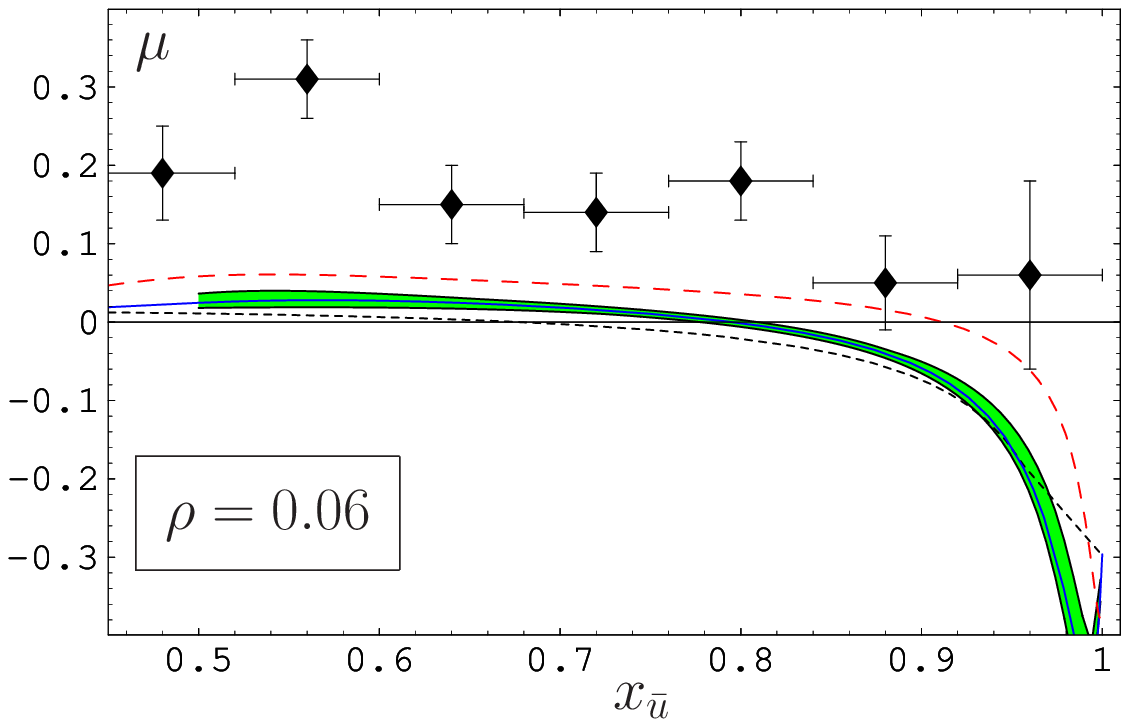}~~%
   \includegraphics[width=0.32\textwidth]{
    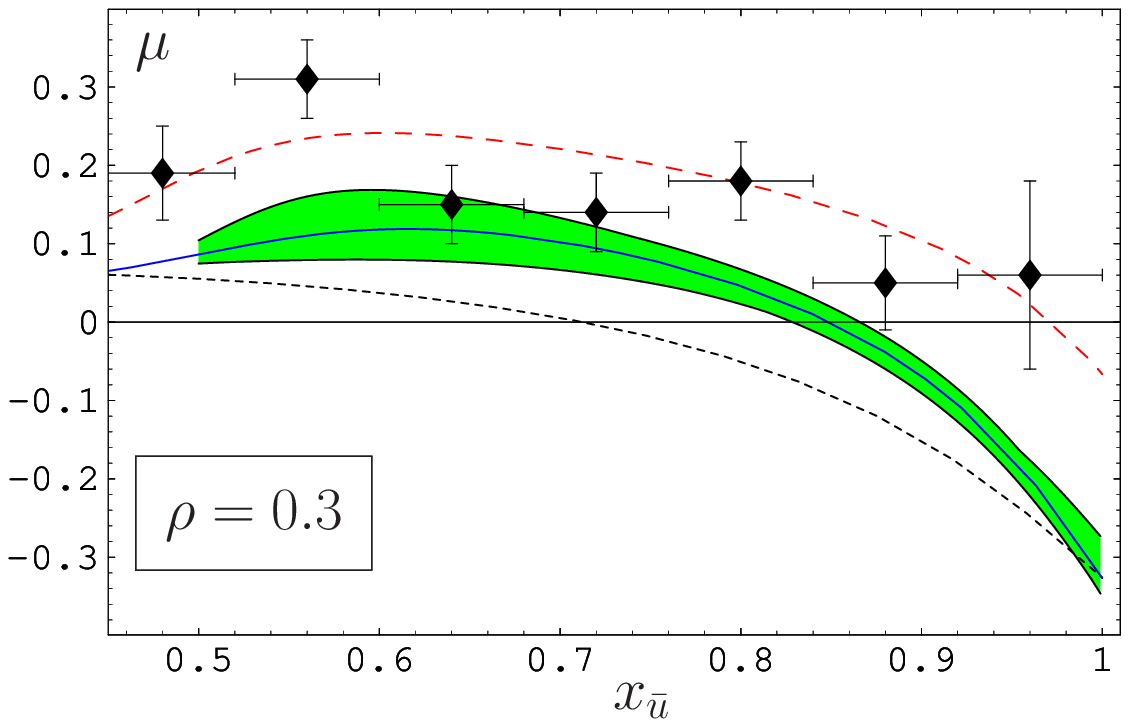}~~%
   \includegraphics[width=0.32\textwidth]{
    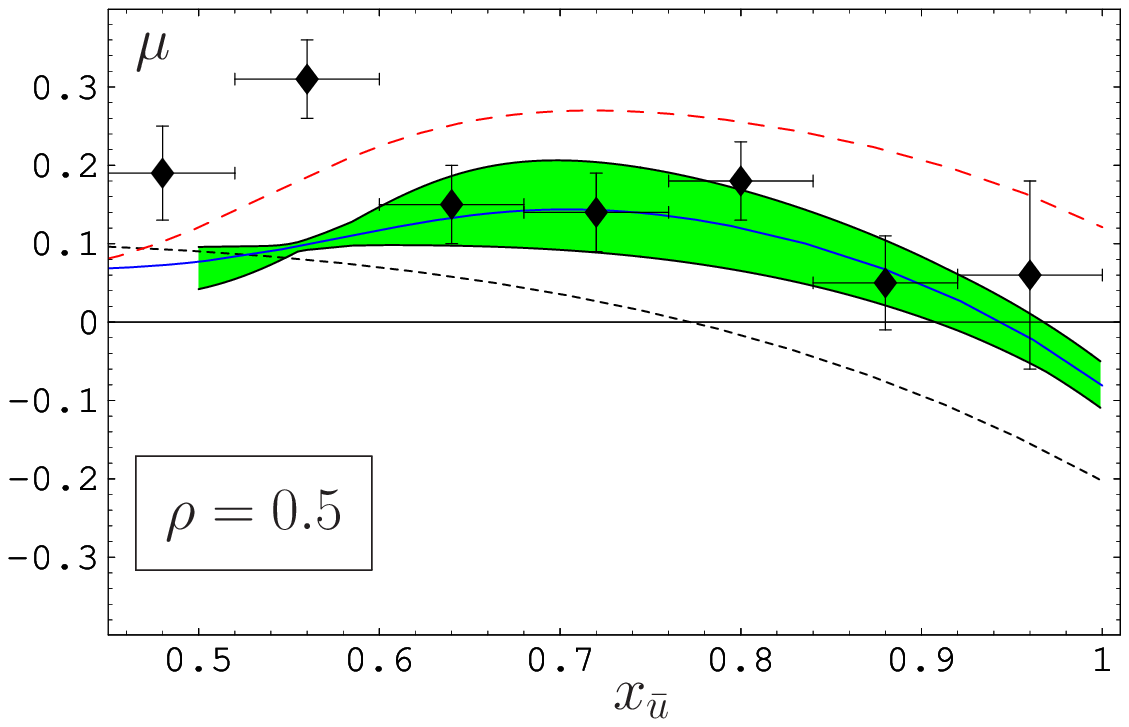}$$
 $$\includegraphics[width=0.32\textwidth]{
    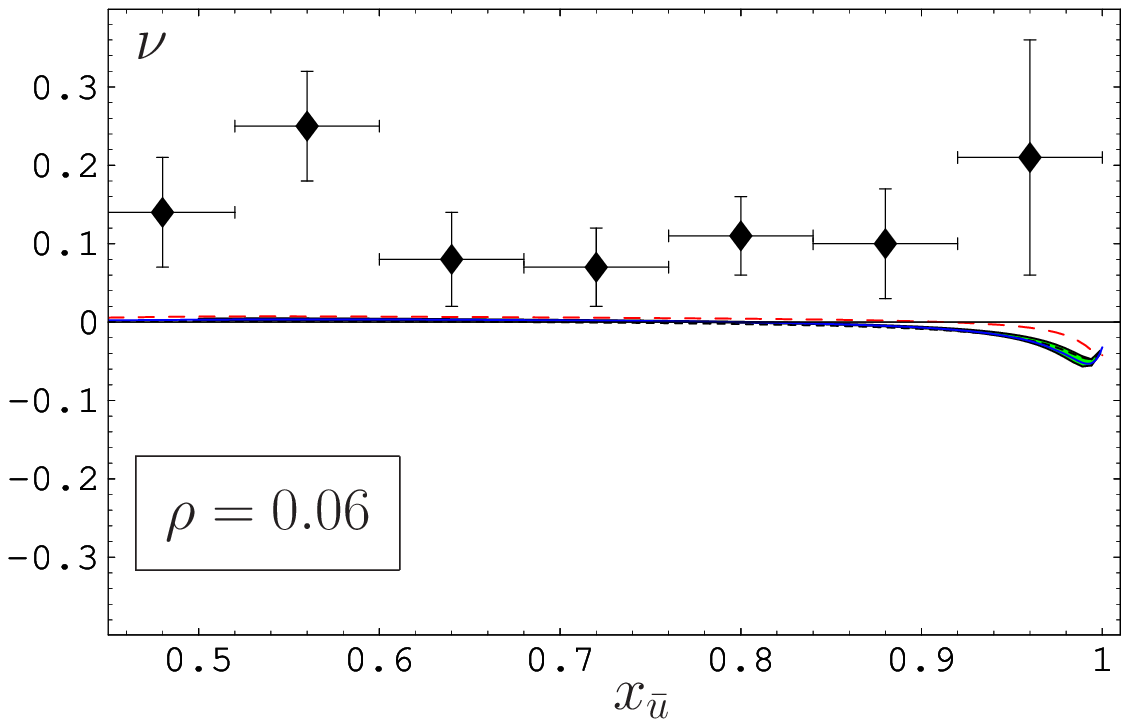}~~%
   \includegraphics[width=0.32\textwidth]{
    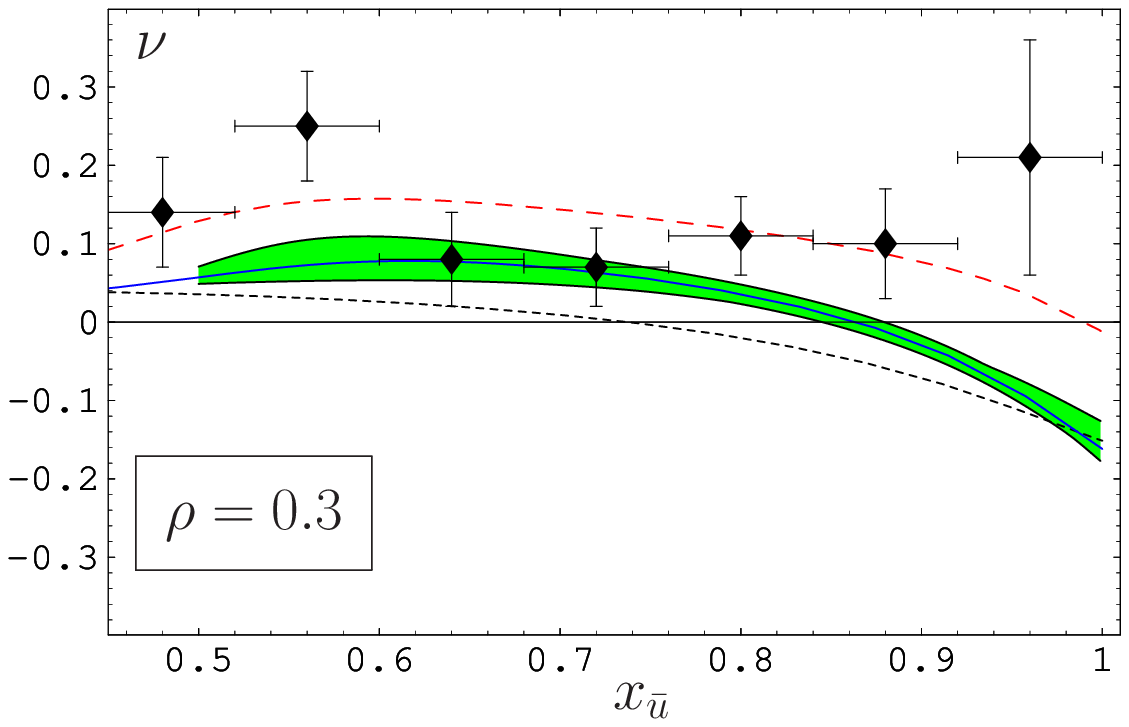}~~%
   \includegraphics[width=0.32\textwidth]{
    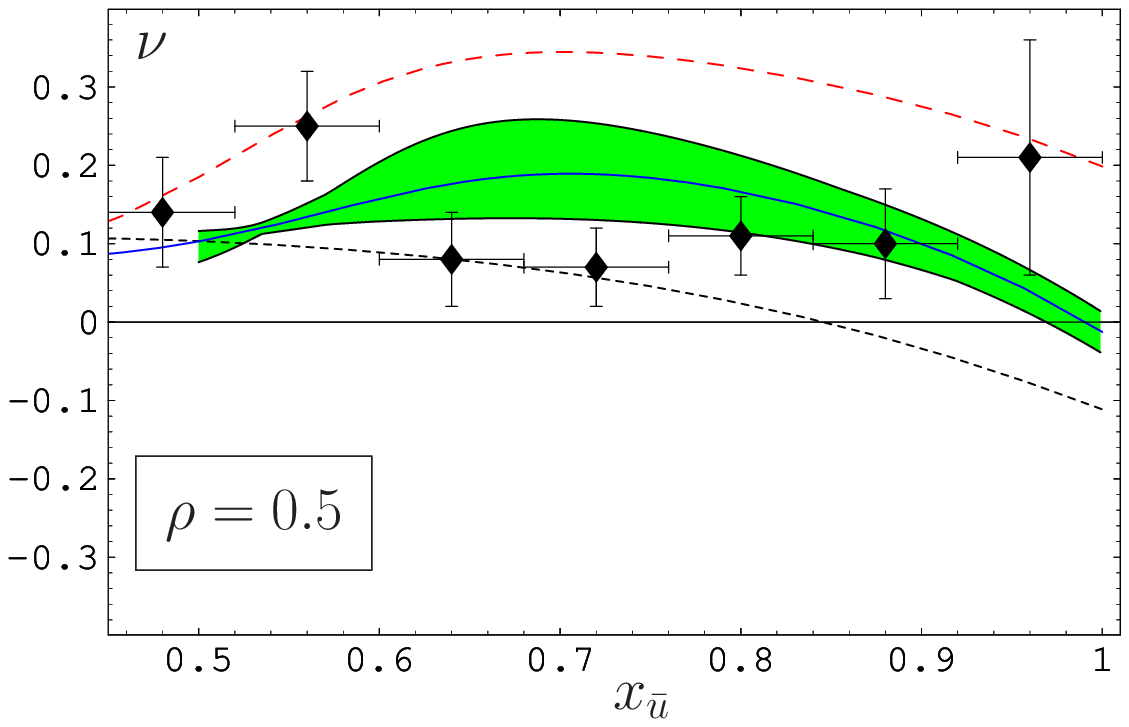}$$
 $$\includegraphics[width=0.32\textwidth]{
    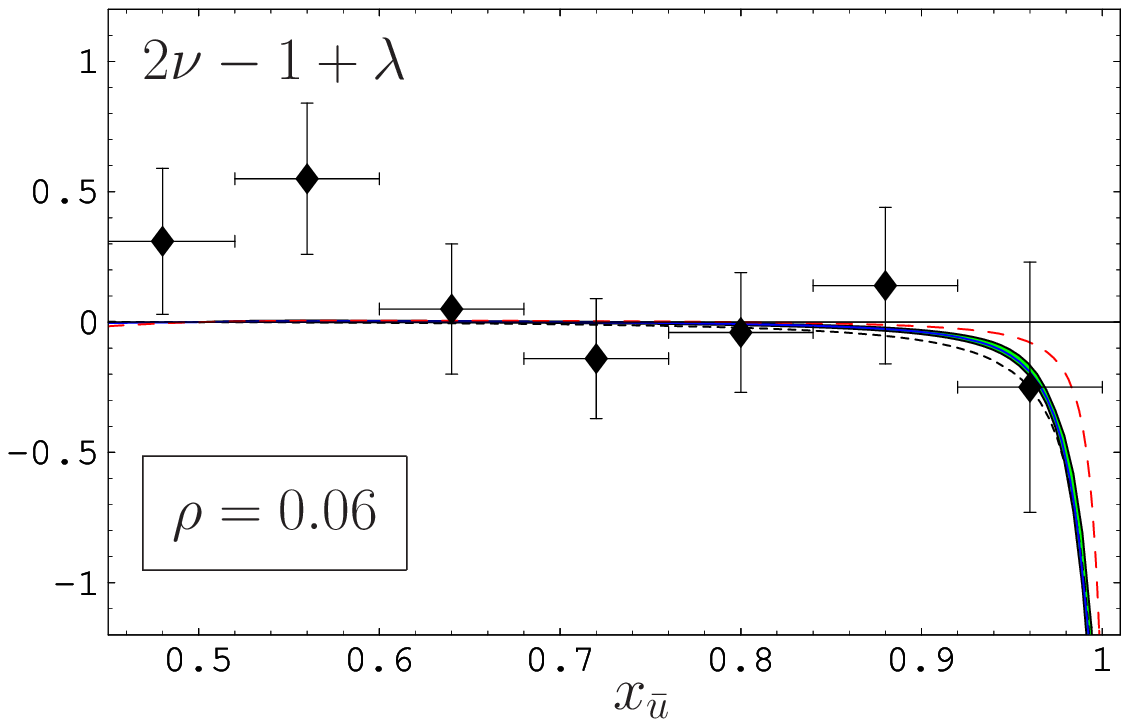}~~%
   \includegraphics[width=0.32\textwidth]{
    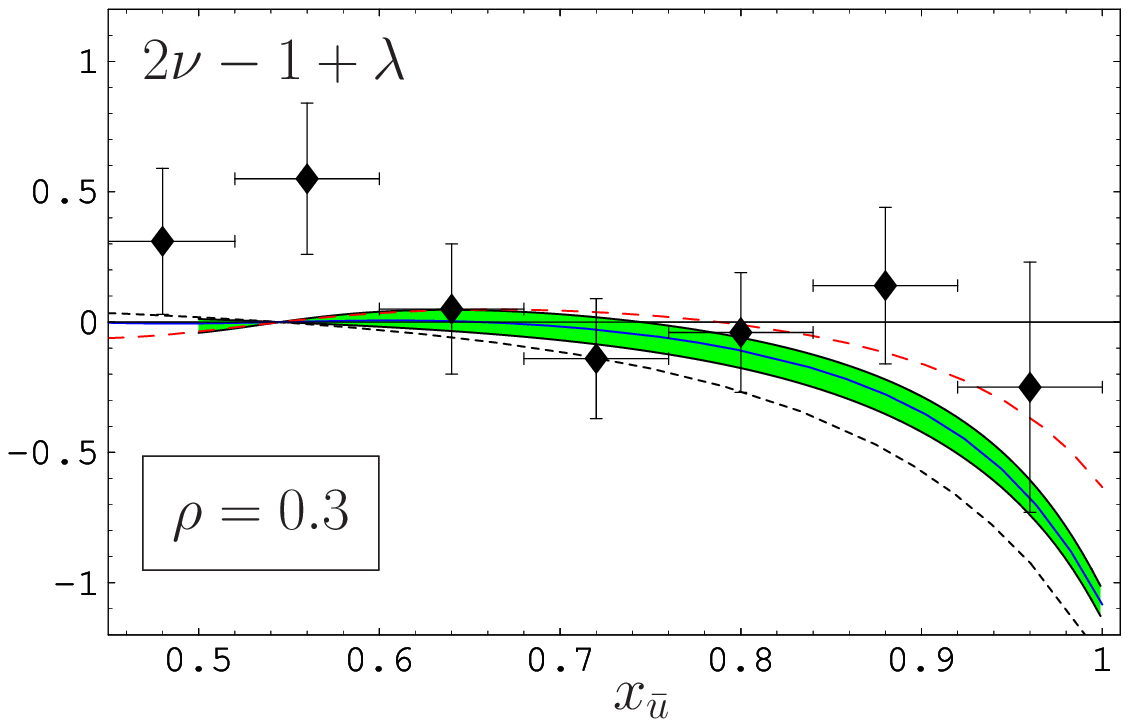}~~%
   \includegraphics[width=0.32\textwidth]{
    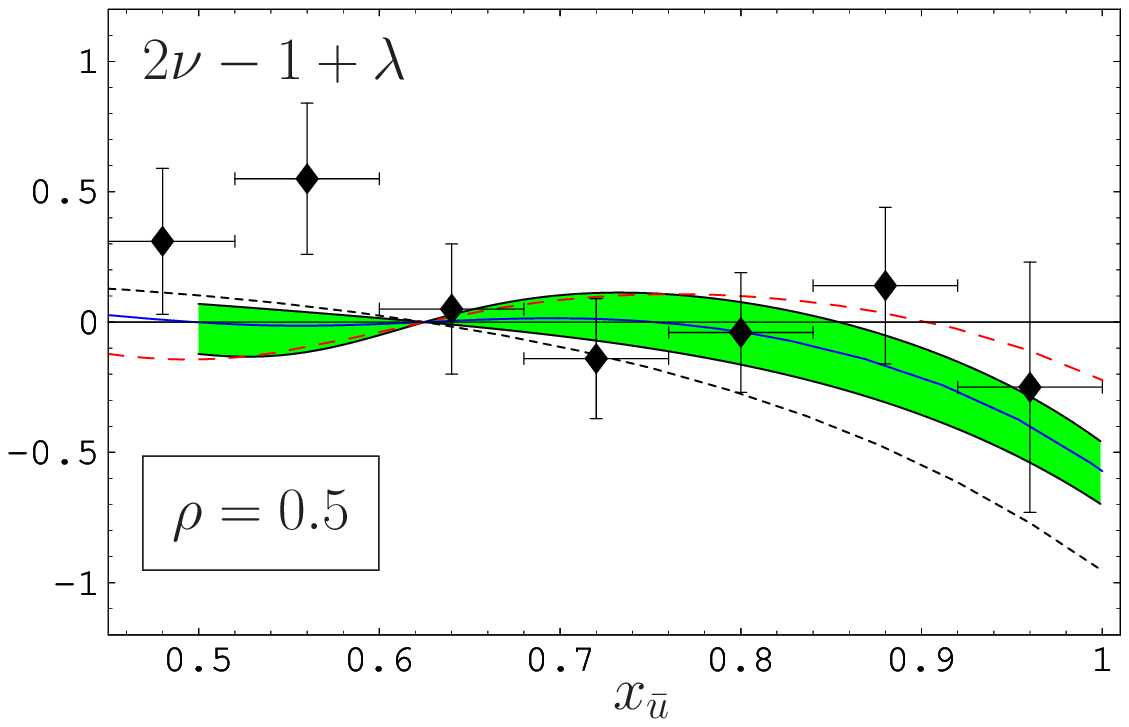}$$
   \vspace{-7mm} \caption{Results for the angular distribution
   parameters $\lambda$, $\mu$, and $\nu$ as functions of
   $x_{\bar{u}}\equiv x_{\pi}$ for different values of
   $\rho\equiv Q_{T}/Q$.
   Predictions for the Lam--Tung \protect\cite{LT80} combination,
   $2\nu -(1-\lambda)$, are also displayed.
   The green strip contains the results for the pion DAs calculated
   with nonlocal QCD sum rules \protect\cite{BMS01,BMS02,BMS03}.
   The (blue) solid line corresponds to one of these
   endpoint-suppressed DAs, termed BMS, while the dotted (black) solid
   line shows the result for the asymptotic DA, and the (red) dashed
   line is the prediction for the endpoint--dominated
   Chernyak--Zhitnitsky DA \protect\cite{CZ84}.
   One-loop evolution of the pion DAs to each maesured $Q^2$ value is
   included.
   The data were taken from \protect\cite{Con89} and were evaluated as
   explained in the text.
   \label{fig:unpol-angular}}
\end{figure}

\begin{figure}[t]
 $$\includegraphics[width=0.30\textwidth]{
    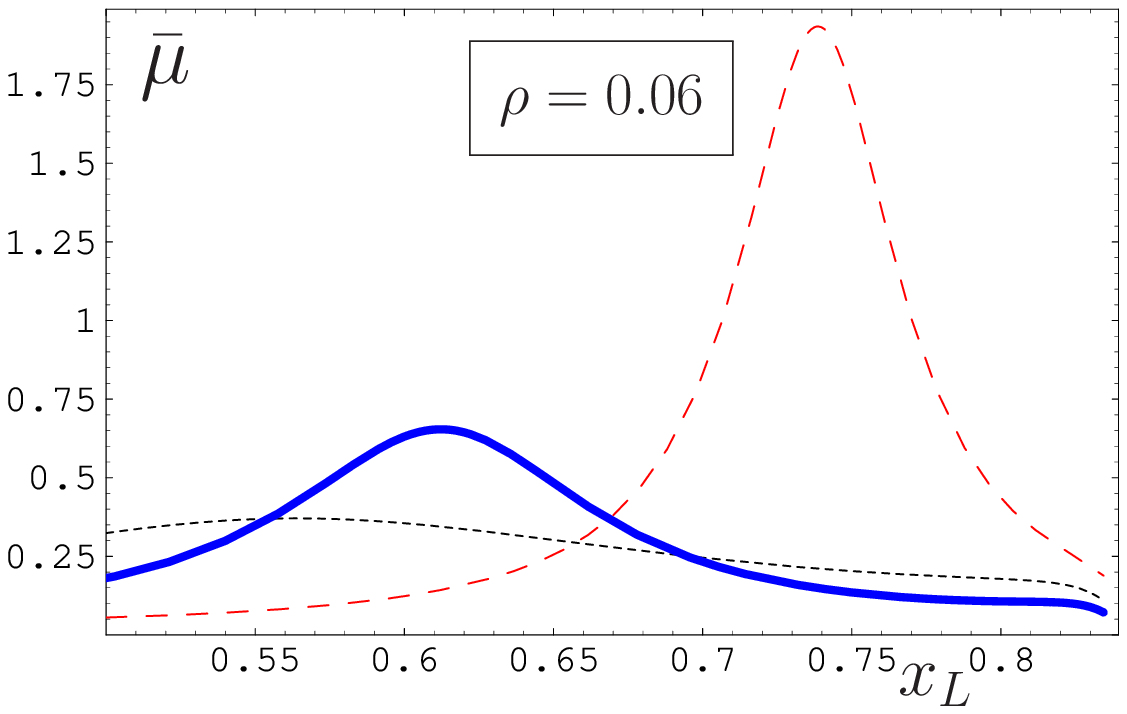}~~%
   \includegraphics[width=0.30\textwidth]{
    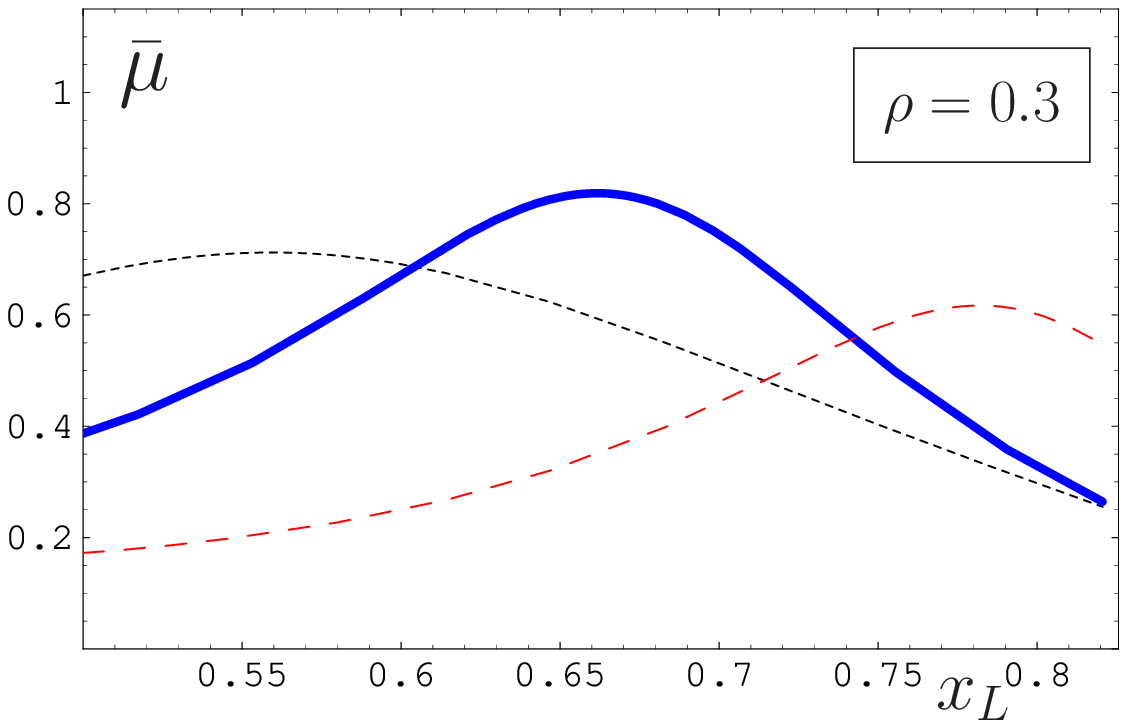}~~%
   \includegraphics[width=0.30\textwidth]{
    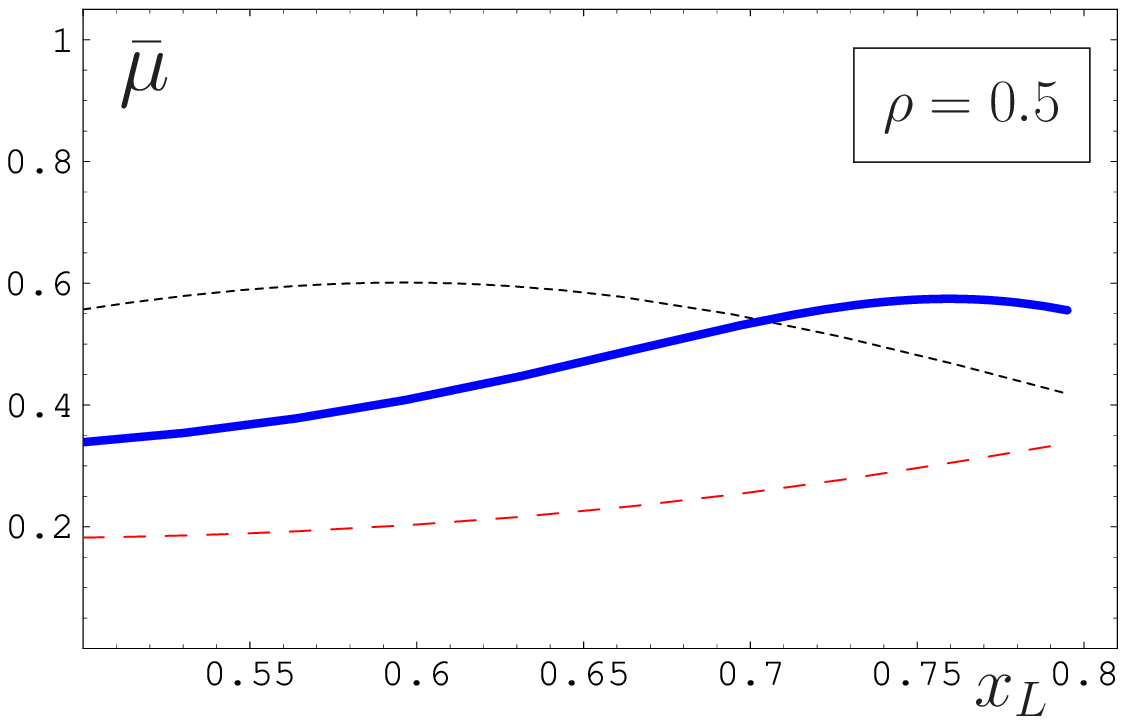}$$
 $$\includegraphics[width=0.30\textwidth]{
    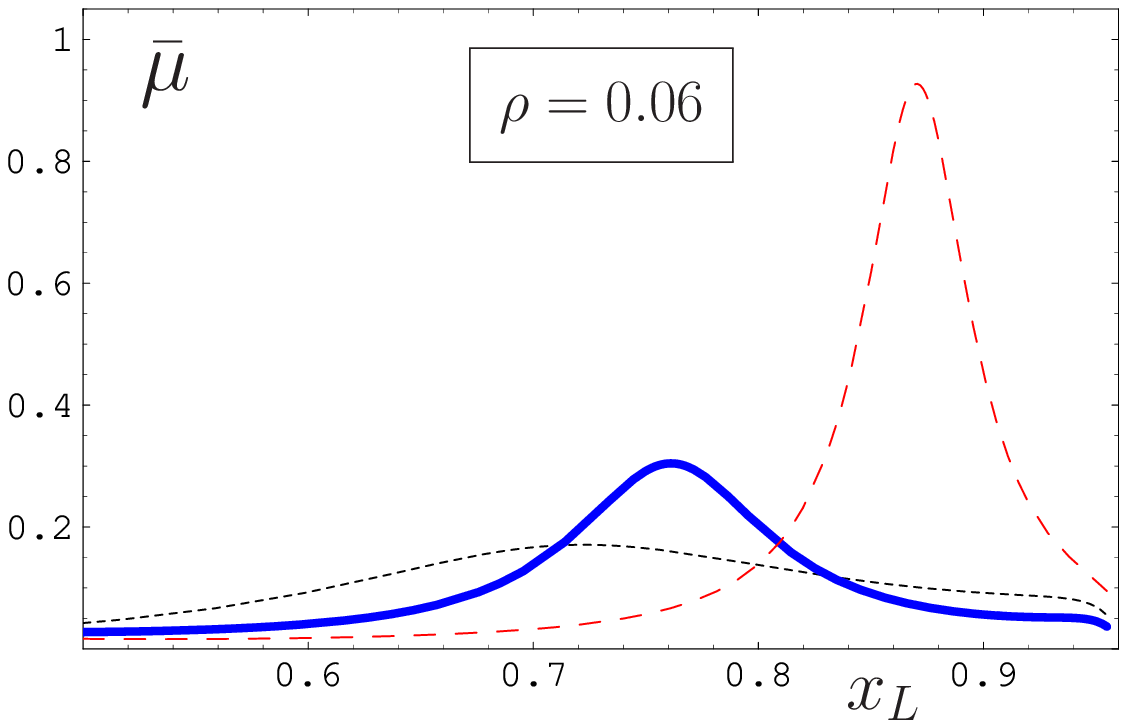}~~%
   \includegraphics[width=0.30\textwidth]{
    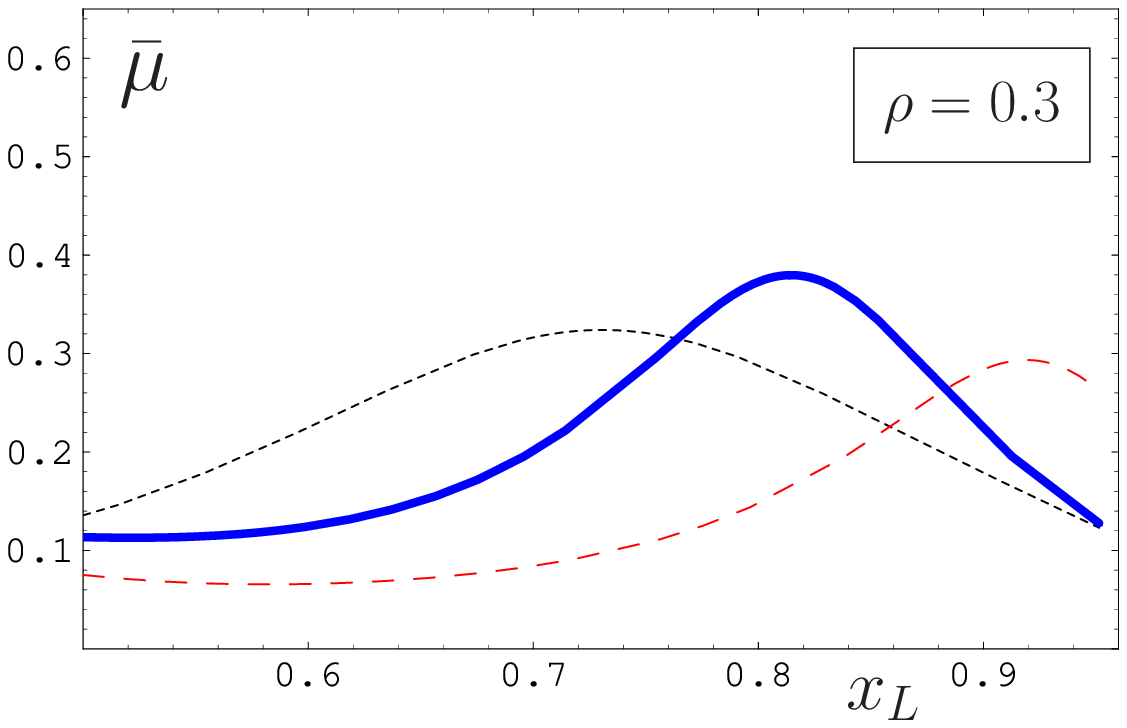}~~%
   \includegraphics[width=0.30\textwidth]{
    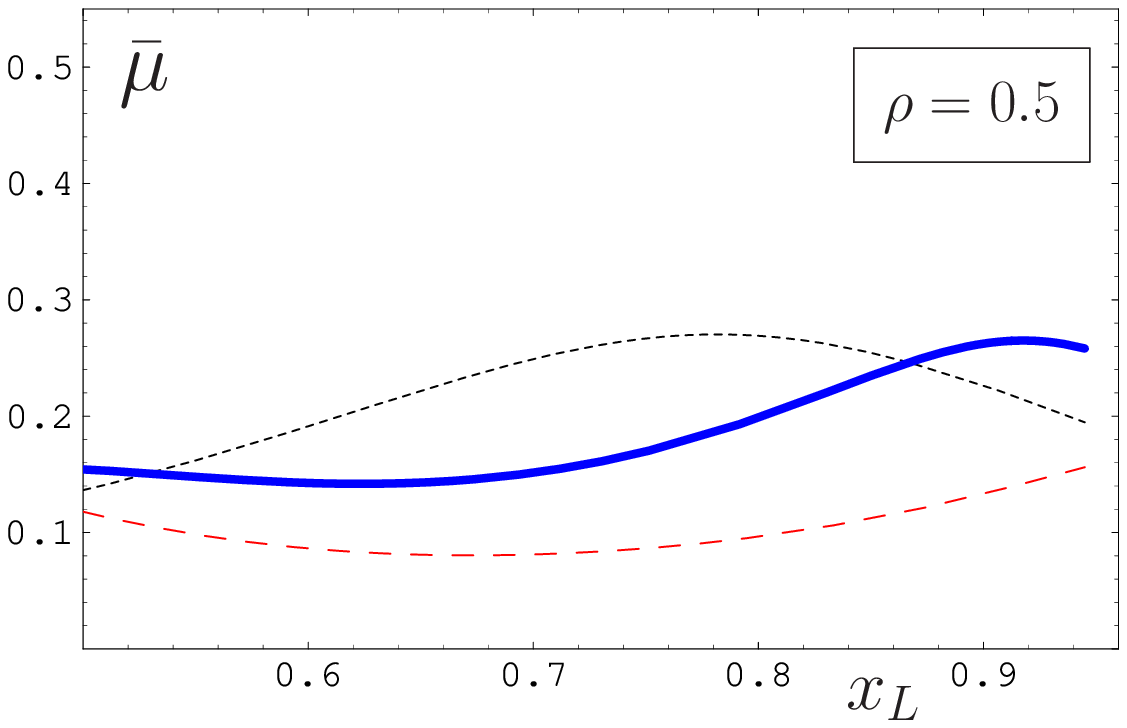}$$
   \vspace{-7mm} \caption{Angular parameter $\bar{\mu}(x_L,\rho)$ as a
   function of $x_{L}$ for three different pion DAs, evaluated at three
   different values of the scaling parameter $\rho$ using one-loop
   evolution.
   Results are shown for the asymptotic pion DA 
   \protect\cite{ER80a,ER80b,LB80} (black dotted line), 
   the BMS model \protect\cite{BMS01} (blue solid line) 
   and the CZ one \protect\cite{CZ84} (red dashed line).
   The upper row corresponds to a center-of-mass energy $s=100$~GeV$^2$,
   whereas the lower one has $s=400$~GeV$^2$.
   \label{fig:rho-evo-1}}\vspace{-2mm}
\end{figure}
\section{Comparison with experimental data}
\label{sec:comp-data}
In this section we present our results for the angular distribution
parameters $\lambda, \mu, \nu$ (versus $x_{\pi}\to x_{\bar{u}}$) for
the unpolarized DY process and compare them with the available
experimental data.
\begin{figure}[th]
 $$\includegraphics[width=0.30\textwidth]{
  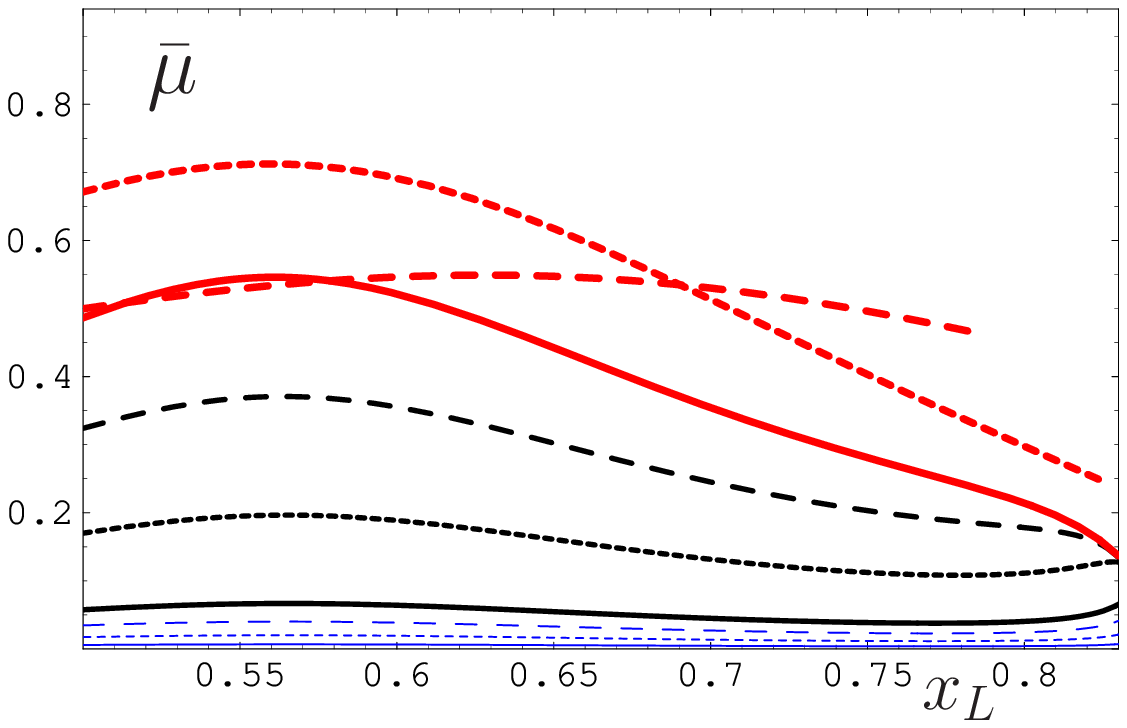}~~%
   \includegraphics[width=0.30\textwidth]{
  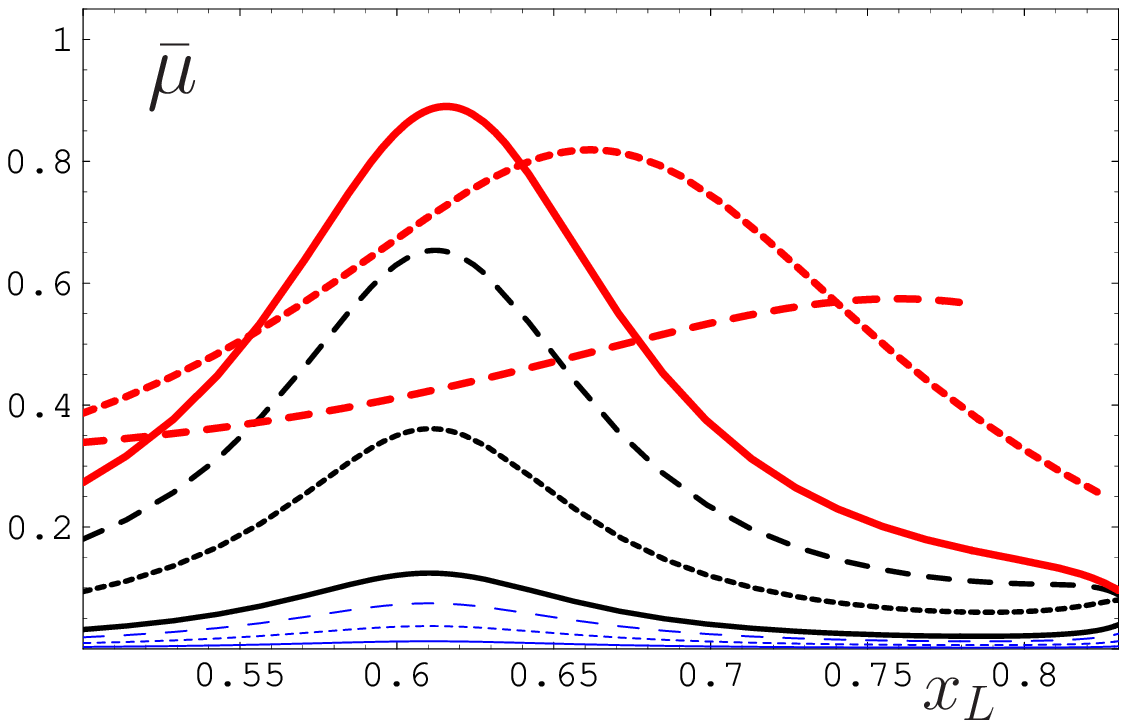}~~%
   \includegraphics[width=0.30\textwidth]{
  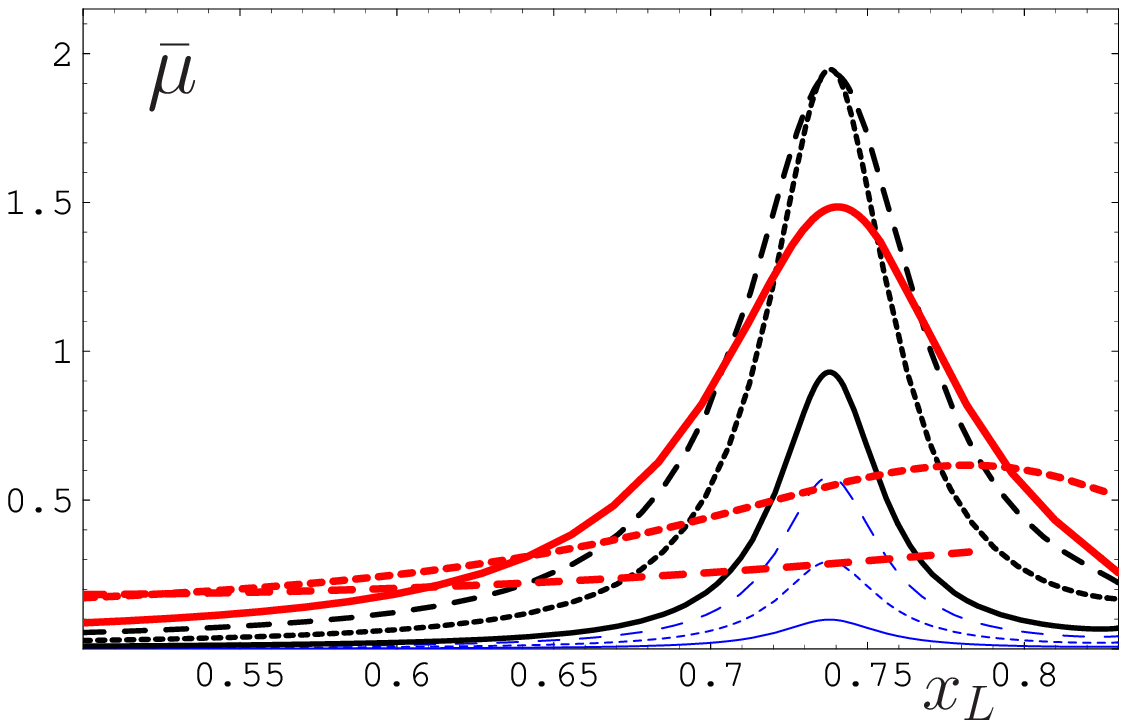}$$
 $$\includegraphics[width=0.30\textwidth]{
  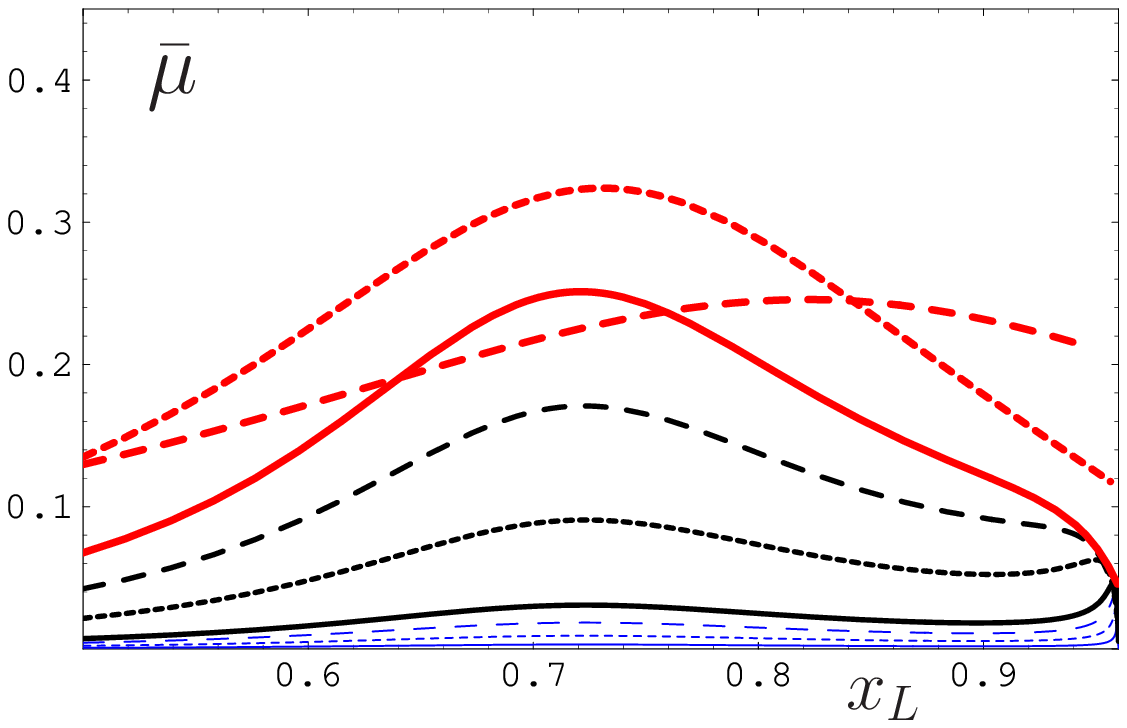}~~%
   \includegraphics[width=0.30\textwidth]{
  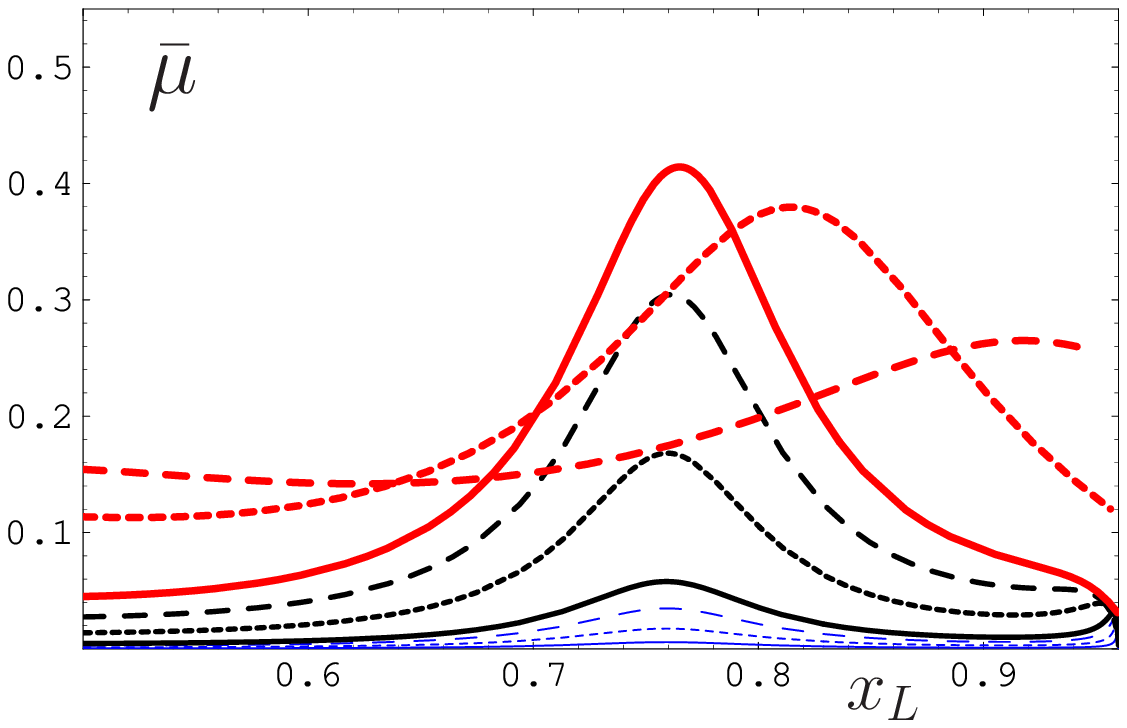}~~%
   \includegraphics[width=0.30\textwidth]{
  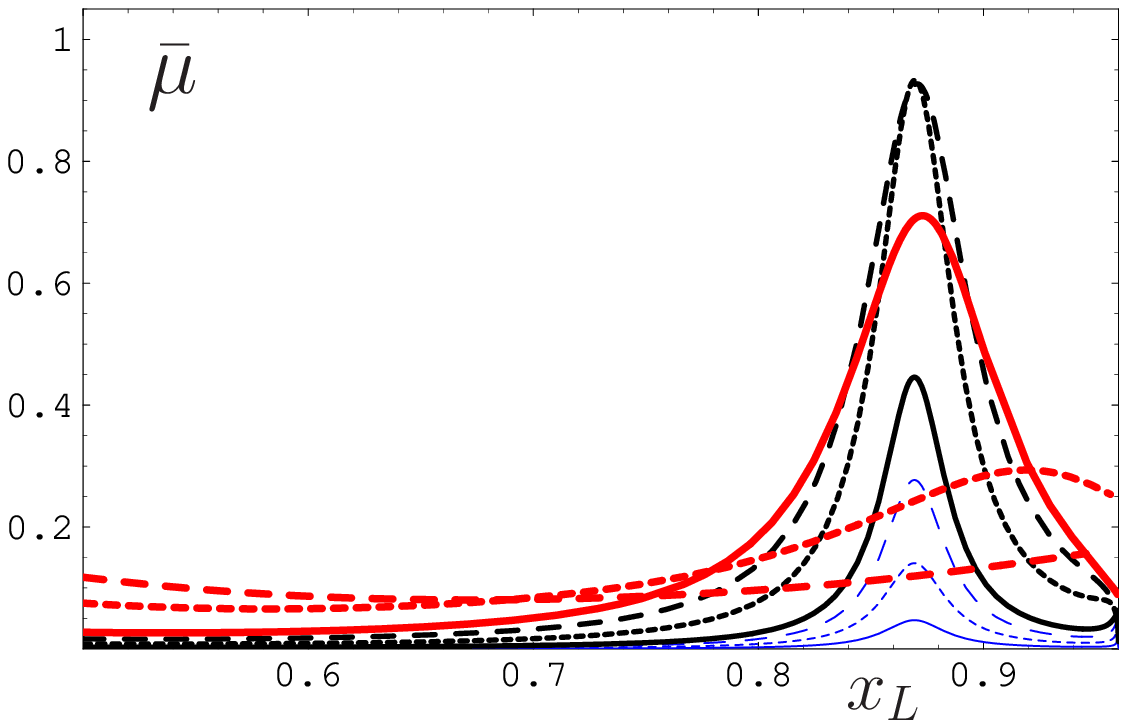}$$
   \vspace{-7mm} \caption{Evolution effect on the angular parameter
    $\bar{\mu}(x_L,\rho)$ vs.\ $x_{L}$ in several steps of the scaling
    parameter $\rho$.
    Results are shown for the asymptotic pion DA 
    \protect\cite{ER80a,ER80b,LB80} (left panel), 
    the BMS model DA \protect\cite{BMS01} (central panel)
    and the CZ one \protect\cite{CZ84} (right panel).
    The following designations are used:
    The solid lines within each color group correspond to the smallest
    $\rho$ value.
    The short-dashed lines denote the results for the intermediate
    $\rho$ values, whereas the long-dashed lines represent the results
    with the largest $\rho$ values.
    The group of the blue lines covers the range $\rho=0.001, 
    0.003, 0.006$;
    the group of black lines gives the results for
    $\rho=0.01, 0.03, 0.06$, whereas the red lines are associated with
    the values $\rho=0.1, 0.3, 0.5$.
    The upper row corresponds to a center-of-mass energy 
    $s=100$~GeV$^2$, whereas the lower one has $s=400$~GeV$^2$.
\label{fig:rho-evo-2}}\vspace{-2mm}
\end{figure}
\begin{figure}[bh]
 $$\includegraphics[width=0.30\textwidth]{
  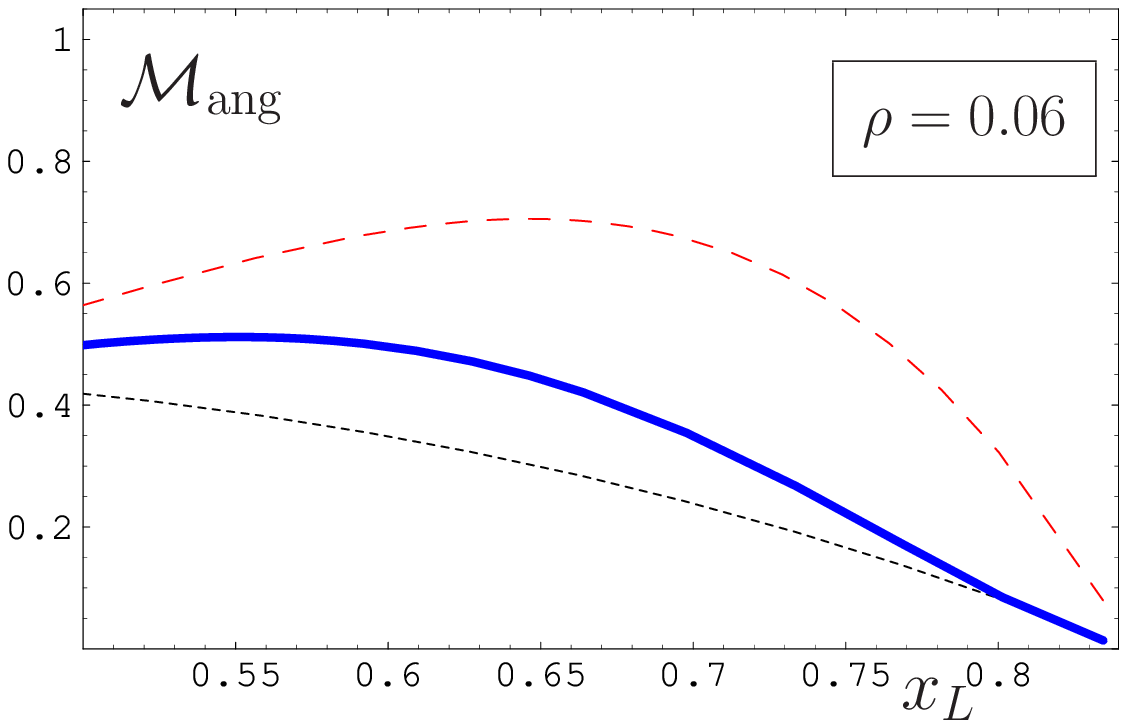}~~%
   \includegraphics[width=0.30\textwidth]{
  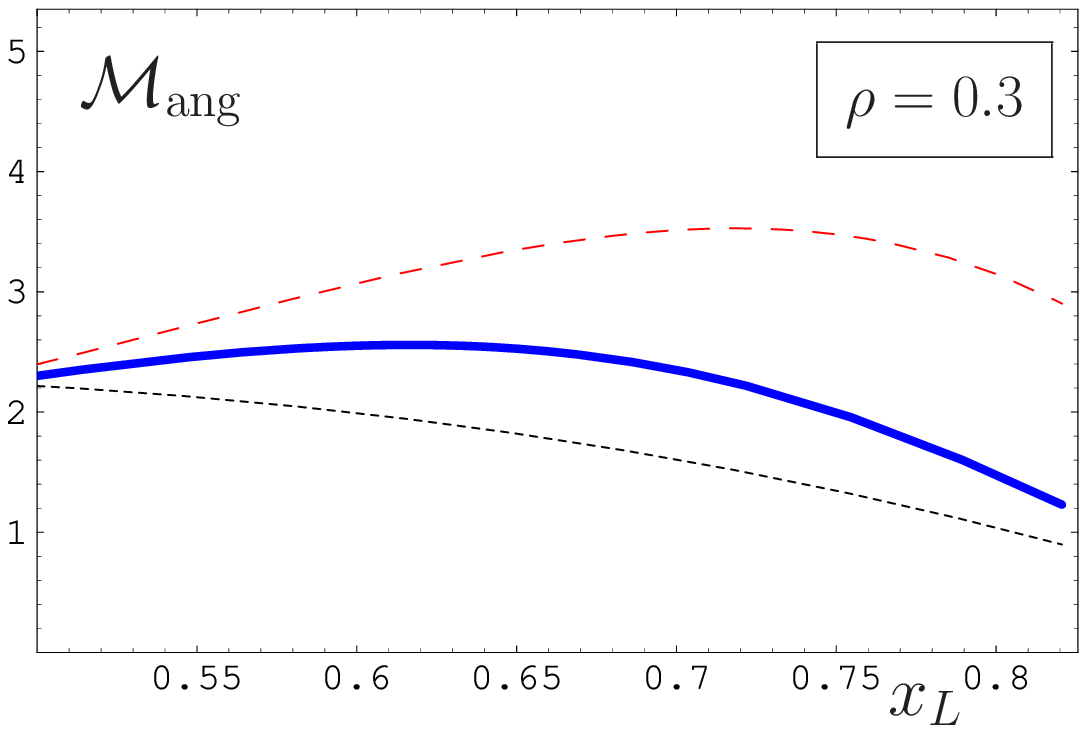}~~%
   \includegraphics[width=0.30\textwidth]{
  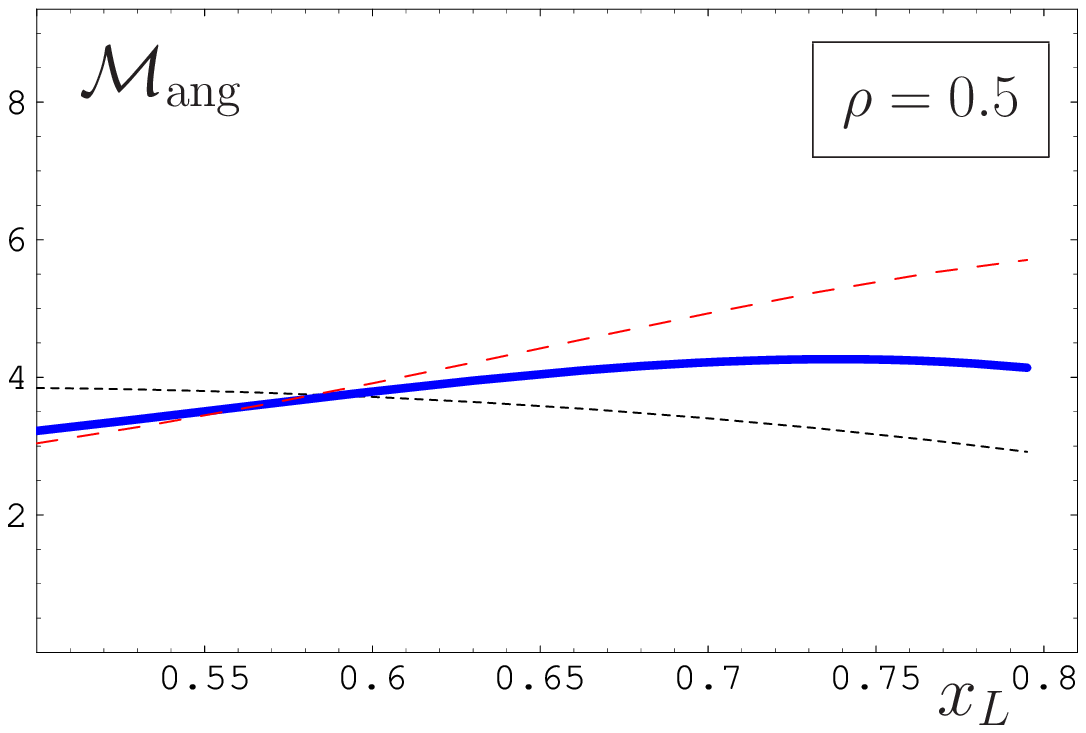}$$
 $$\includegraphics[width=0.30\textwidth]{
  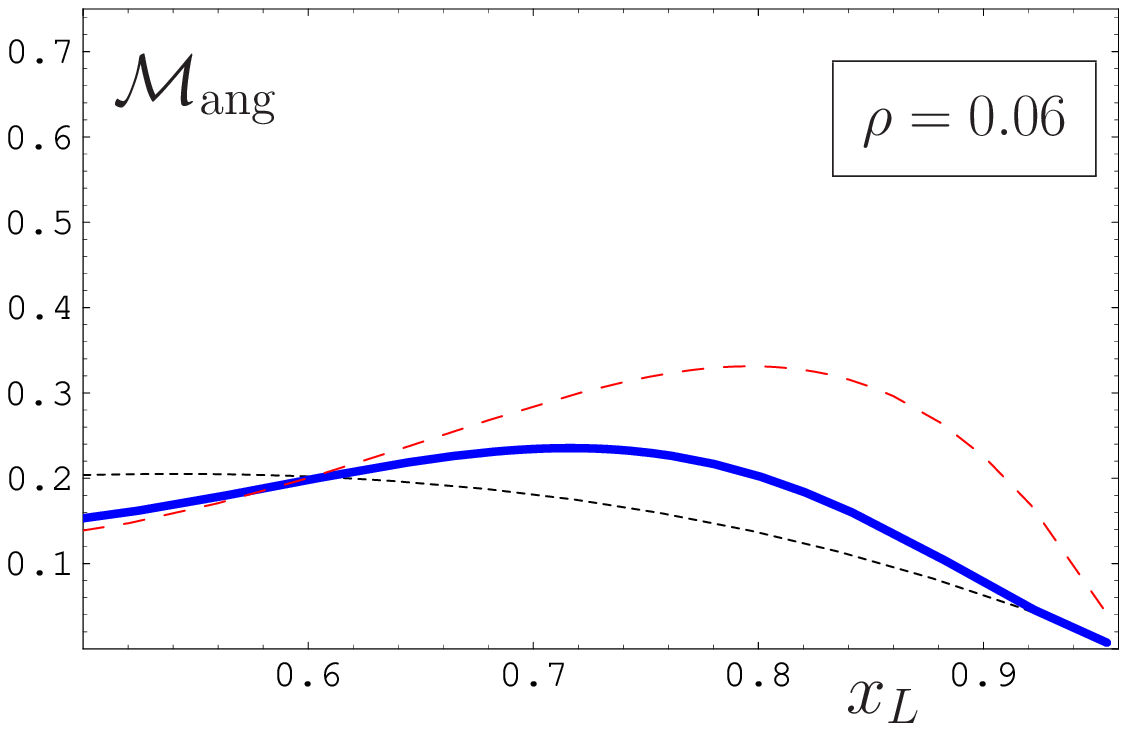}~~%
   \includegraphics[width=0.30\textwidth]{
  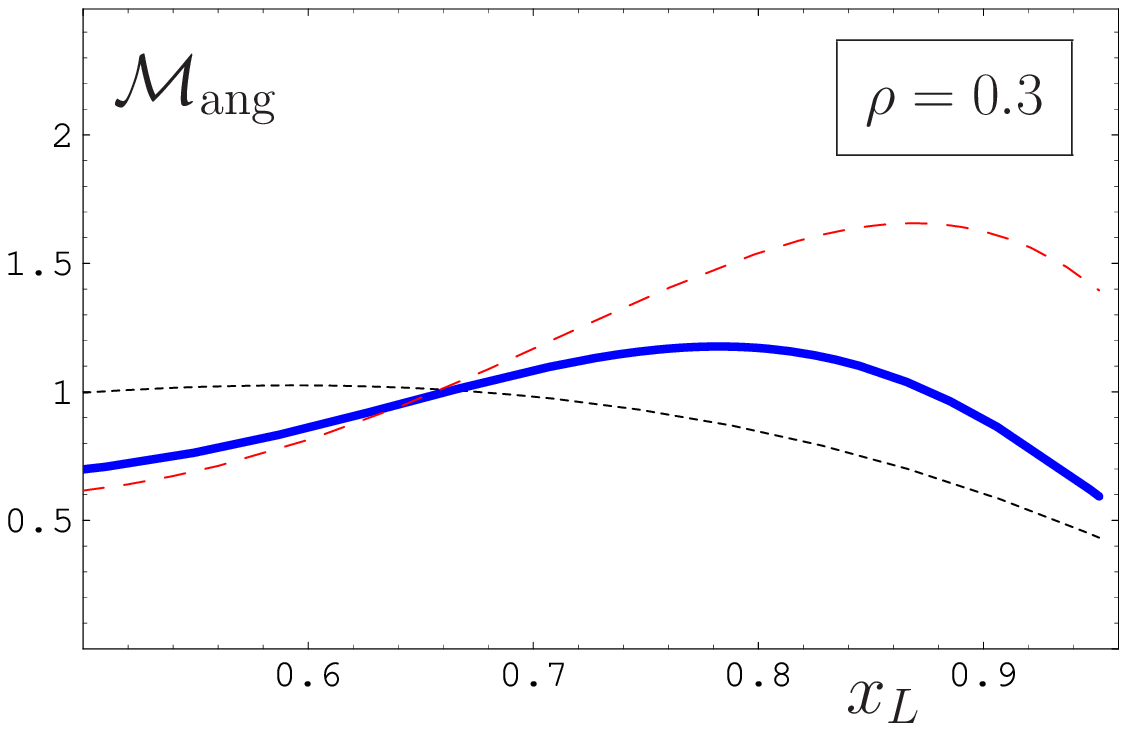}~~%
   \includegraphics[width=0.30\textwidth]{
  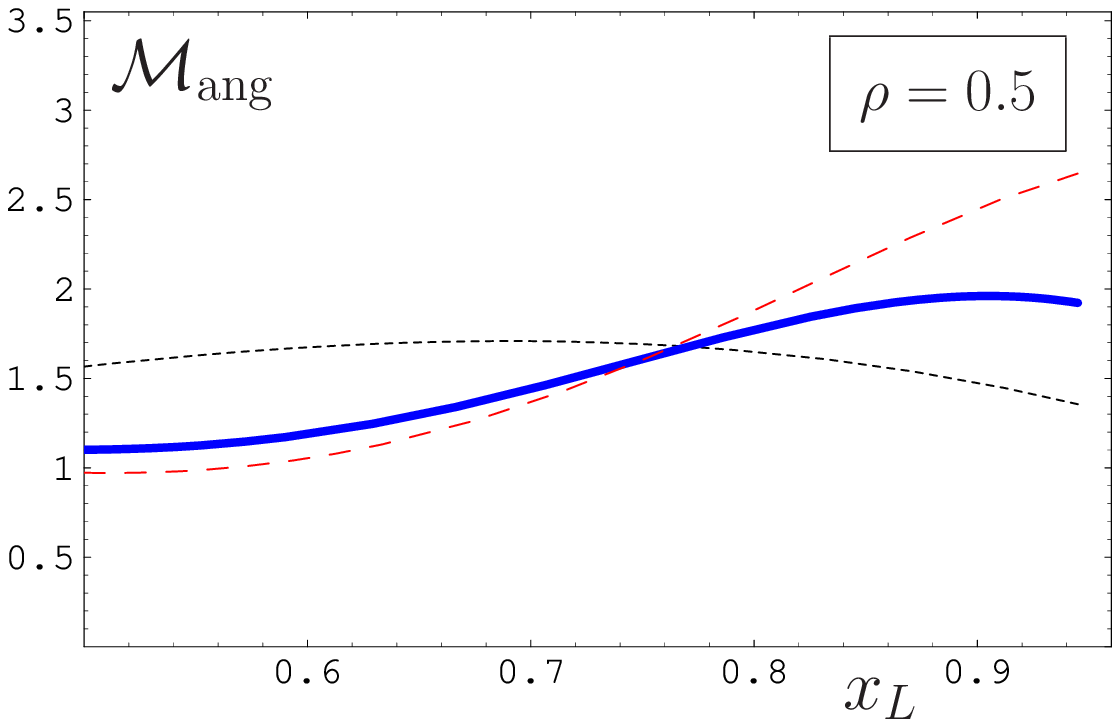}$$
   \vspace{-7mm} \caption{Angular moment ${\cal M}_\text{ang} $ as a
   function of $x_{L}$ for three different pion DAs, evaluated at three
   different values of the scaling parameter $\rho$ using one-loop
   evolution.
   Results are shown for the asymptotic pion DA 
   \protect\cite{ER80a,ER80b,LB80} (black dotted line), 
   the BMS model \protect\cite{BMS01} (blue solid
   line) and the CZ one \protect\cite{CZ84} (red dashed line).
   The upper row corresponds to a center-of-mass energy $s=100$~GeV$^2$,
   whereas the lower one has $s=400$~GeV$^2$.
   \label{fig:rho-evo-Ang-Mom}}
\end{figure}
We also include predictions for the parameter $\bar{\mu}$ (versus
$x_{L}$), which is nonzero only in the polarized DY process, and the
SSA $\cal A$.

All results were obtained using different models for the pion DA,
already mentioned, and they are shown at different values of $Q_{T}$,
i.e., at different values of the scaling parameter $\rho$.

To be specific, we compare in Fig.\ \ref{fig:unpol-angular} our
theoretical predictions for $\lambda, \mu, \nu$ with experimental
data from E615 \cite{Con89} in that $x_\pi$ region reported by this
collaboration using a $252$~GeV $\pi^{-}$ beam interacting in
an unpolarized tungsten target.

We used for convenience the Gottfried--Jackson frame and included in
our analysis the data sample listed in their Table VIII.
Because the range of the probed transverse photon momentum, $Q_{T}$,
is in our opinion too large for averaging, we followed another
strategy than the authors in Ref.\ \cite{BBKM94}.
Notably, we adopted some value of the scaling parameter $\rho=Q_{T}/Q$
and required the momenta $Q_{T}$ and $Q$ to be within the reported
window of the measurement in \cite{Con89}.
We evaluated this way the angular parameters shown in the figures,
obtained with different pion DAs, including one-loop ERBL evolution
with $Q^2=m_{\mu\mu}^2$, and for three different values of
$\rho=0.06, 0.3, 0.5$, whereas $s=500$~GeV${}^2$ in accordance with
\cite{Con89}.
Strictly speaking, the results at too low values of $\rho\lesssim 0.06$
are, actually, not compatible with factorization because then $Q_{T}$
becomes of the order of $\Lambda_{\rm QCD}$.
They are shown here merely for illustration purposes.

One important observation from Fig.\ \ref{fig:unpol-angular} is that
the parameter $\mu$ increases proportionally with $\rho$, as it is
qualitatively expected from Eq.\ (\ref{eq:mu.unp}) (though there is
an additional term proportional to $\rho^{3}$, absent in the parton
DY model).
This increase turns out to be small for the asymptotic DA 
\cite{ER80a,ER80b,LB80}, whereas it becomes rather too strong for the 
CZ model \cite{CZ84}, while for the BMS strip \cite{BMS01} this 
enhancement with $\rho$ is moderate and provides best agreement with 
the data.
A similar behavior is seen also for $\nu$ in this figure, which,
according to Eq.\ (\ref{eq:nu.unp}), should increase proportionally to
$\rho^{2}$.
Also here the BMS strip compares most favorably with the data relative
to the other options.
On the other hand, the Lam--Tung combination $2\nu-1+\lambda$ 
\cite{LT80}, which is the analogue of the Callan--Gross relation in 
deeply inelastic scattering (and also reflects the kinematical nature 
of the azimuthal asymmetry~\cite{OT05}), is badly violated by the data.

This trend is in agreement with the theoretical predictions above
approximately $x_{\bar{u}}\sim 0.6$, though, at very large 
$x_{\bar{u}}$ close to the kinematic limit, all tested model DAs 
tend to fall stronger than the data.

\begin{figure}[b]
 $$\includegraphics[width=0.30\textwidth]{
 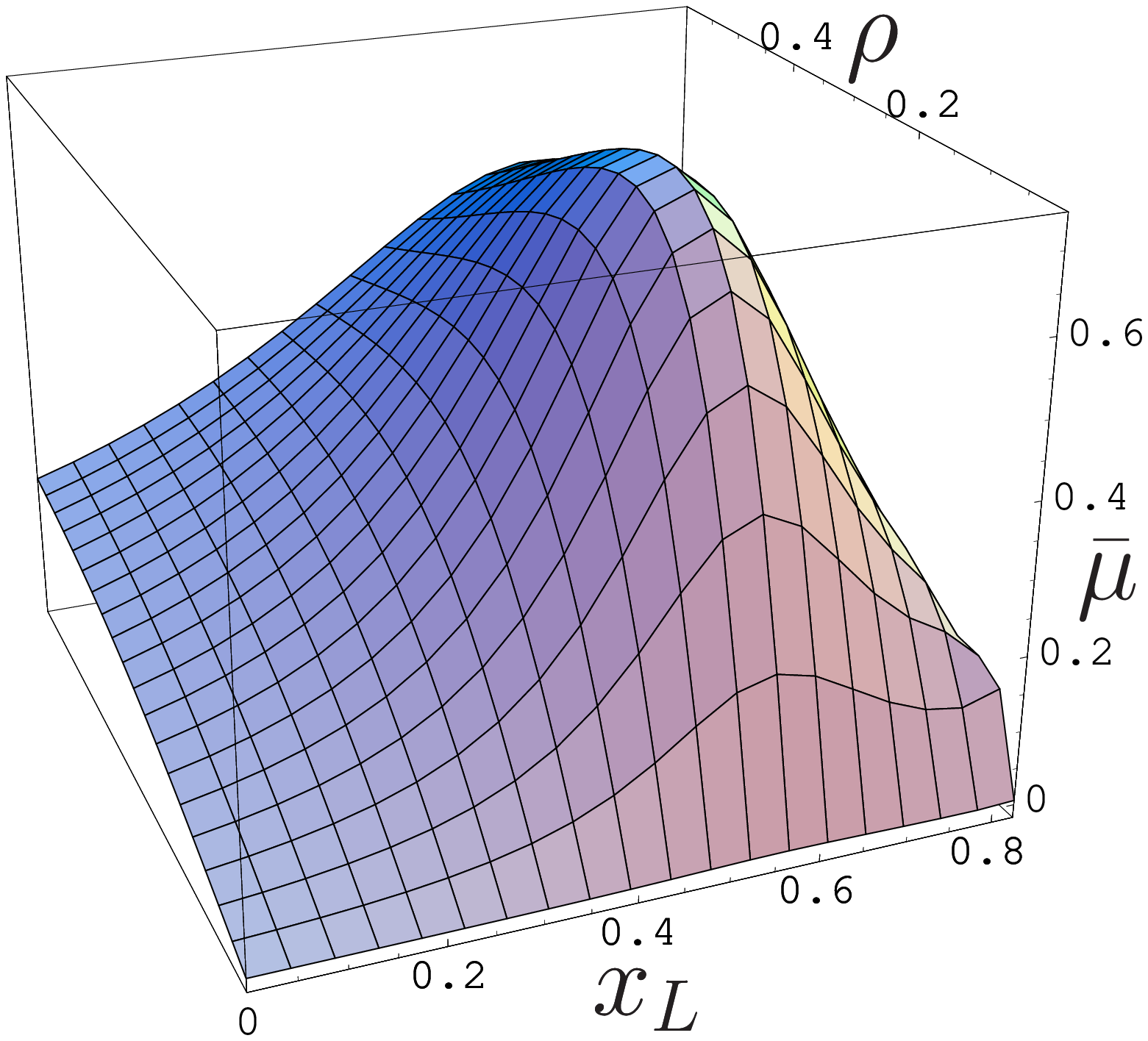}~~%
   \includegraphics[width=0.30\textwidth]{
 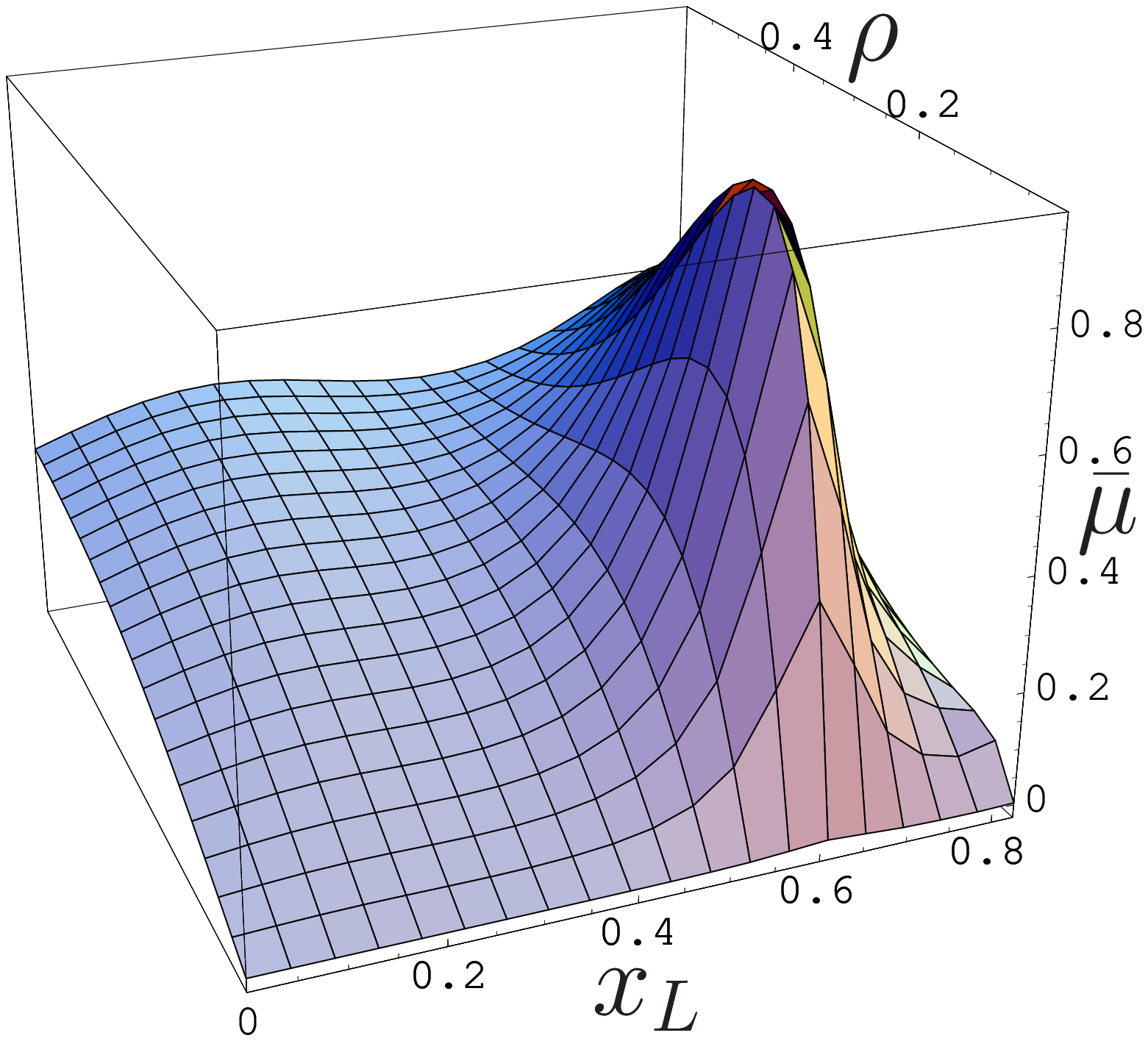}~~%
   \includegraphics[width=0.30\textwidth]{
 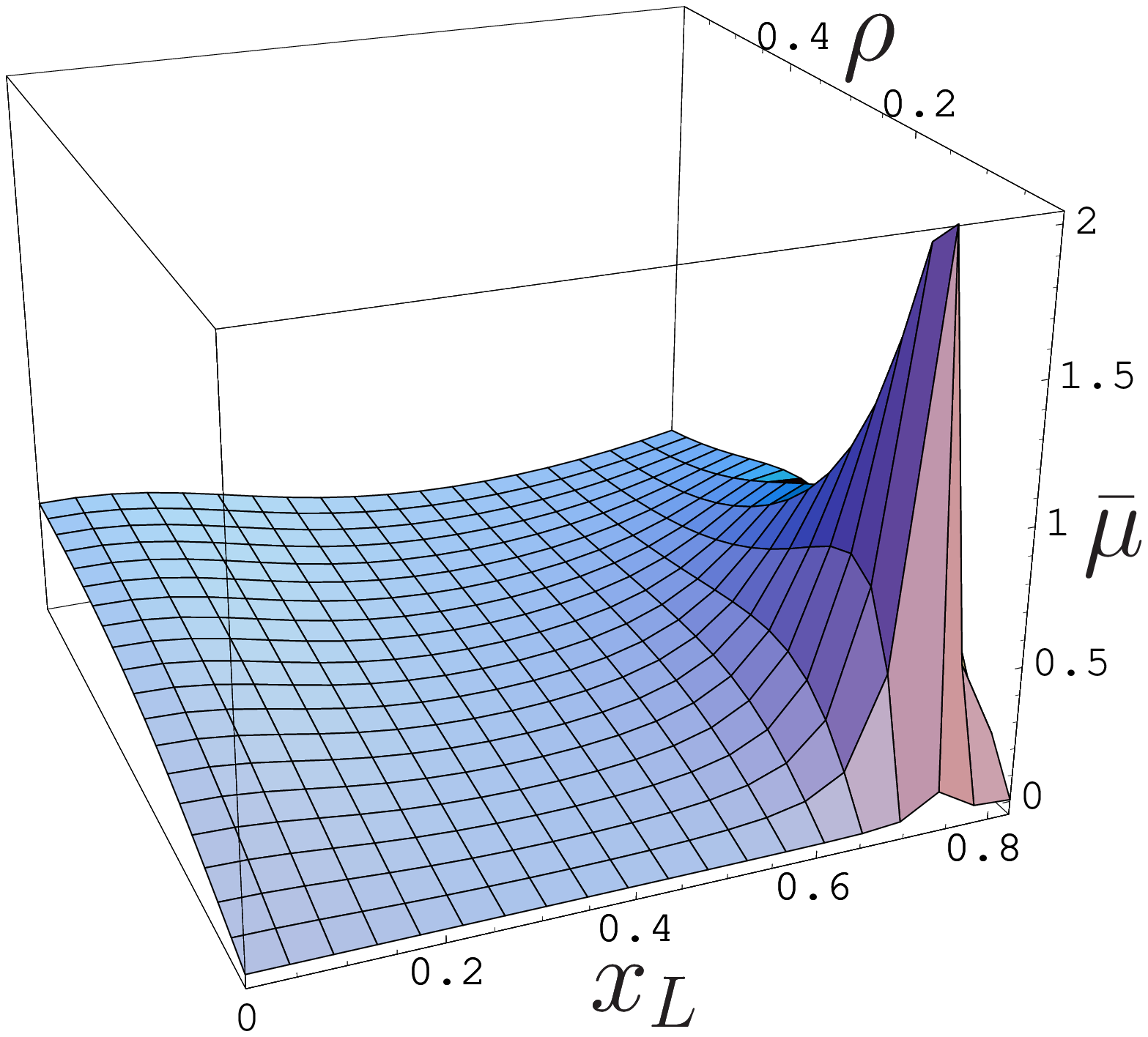}$$
 $$\includegraphics[width=0.30\textwidth]{
 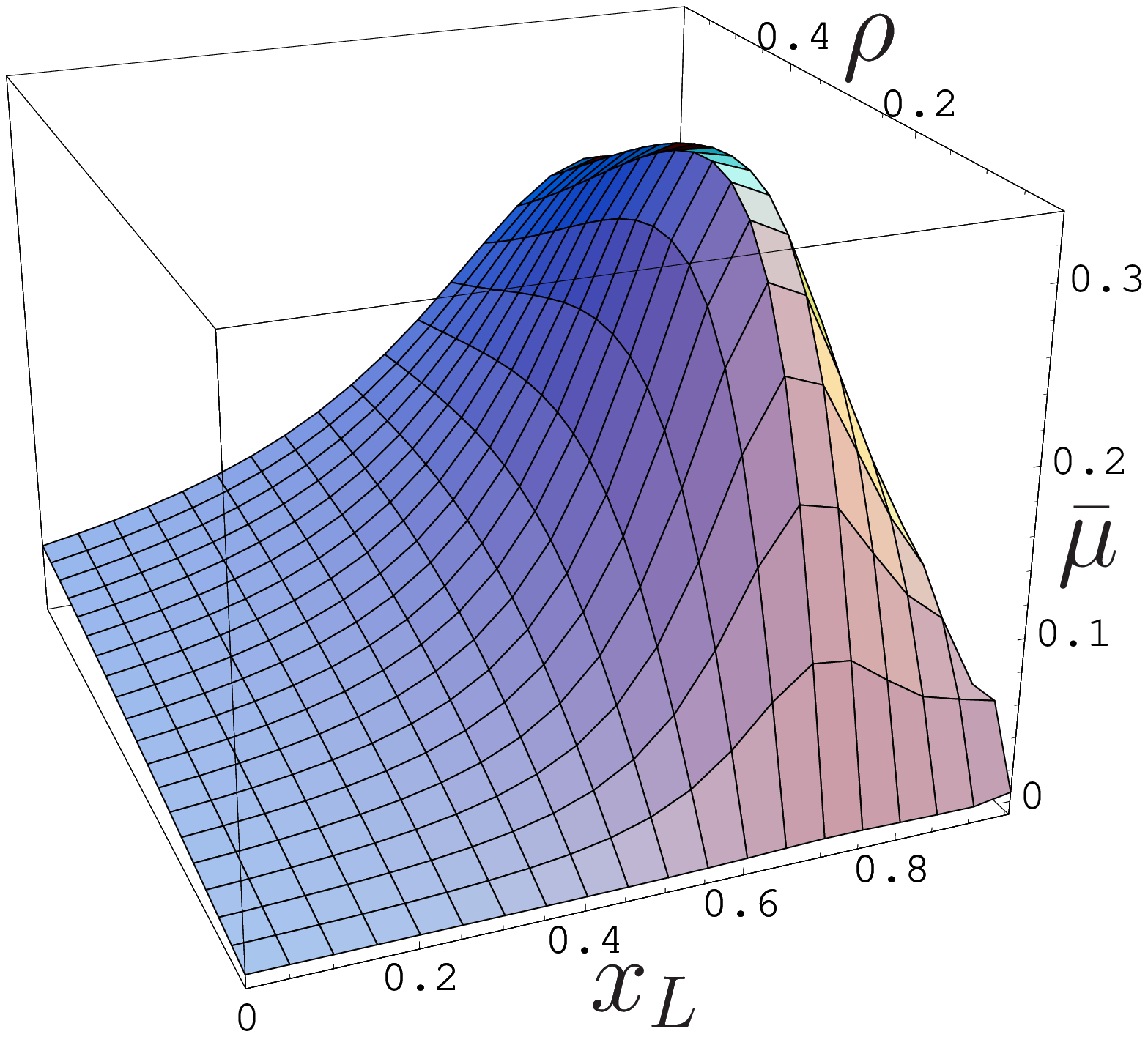}~~%
   \includegraphics[width=0.30\textwidth]{
 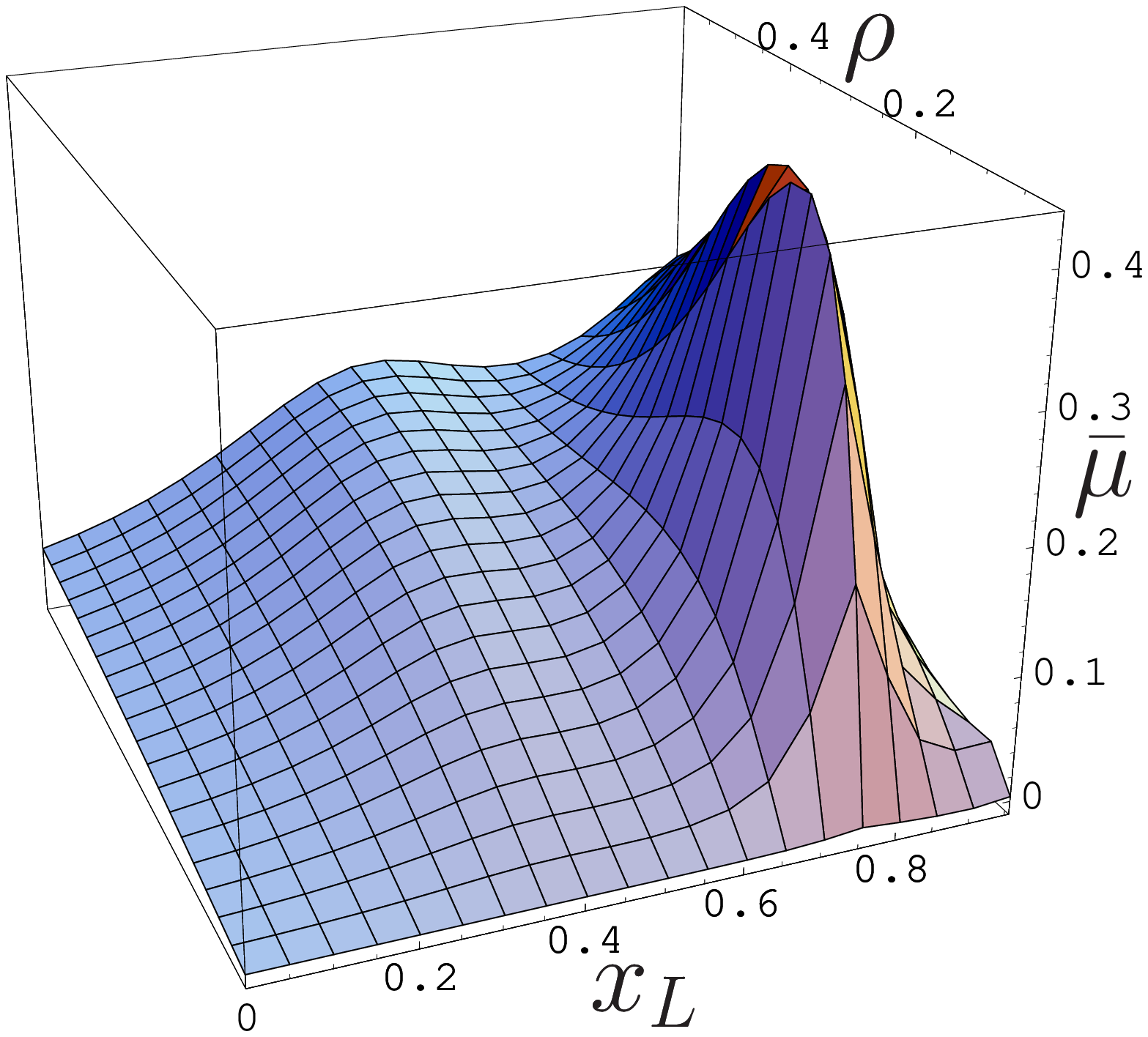}~~%
   \includegraphics[width=0.30\textwidth]{
 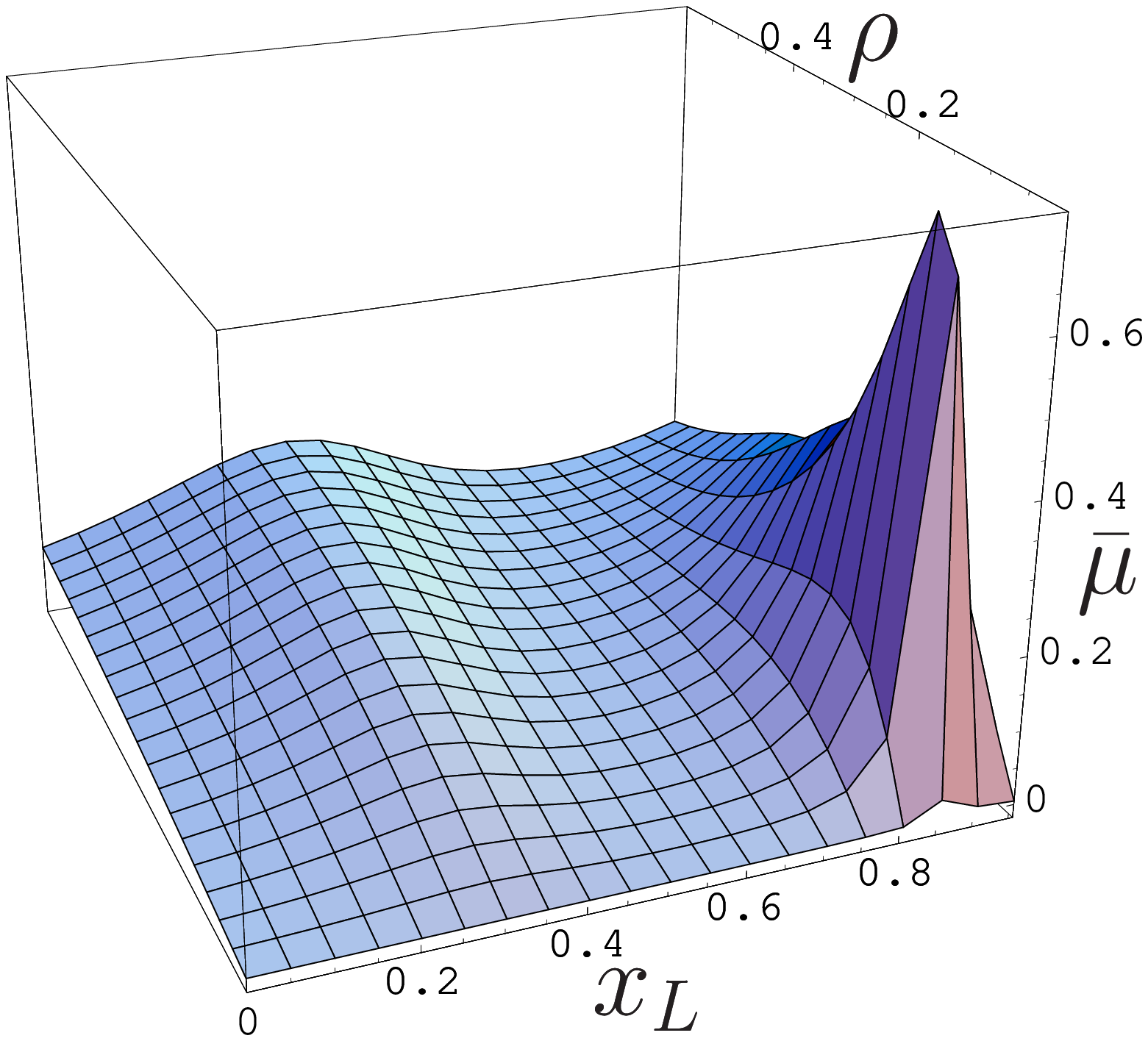}$$
   \vspace*{-7mm} \caption{3D-plots of the angular parameter
    $\bar{\mu}(x_L,\rho)$ for three different choices of the pion DA:
    Asymptotic \protect\cite{ER80a,ER80b,LB80} (left), BMS model
    \protect\cite{BMS01} (center), and CZ model \protect\cite{CZ84}
    (right).
    The upper row corresponds to a center-of-mass energy $s=100$~GeV$^2$,
    whereas the lower one has $s=400$~GeV$^2$.
   \label{fig:rho-evo-2-3D}}\vspace*{-2mm}
\end{figure}
\begin{figure}[th]
 $$\includegraphics[width=0.30\textwidth]{
  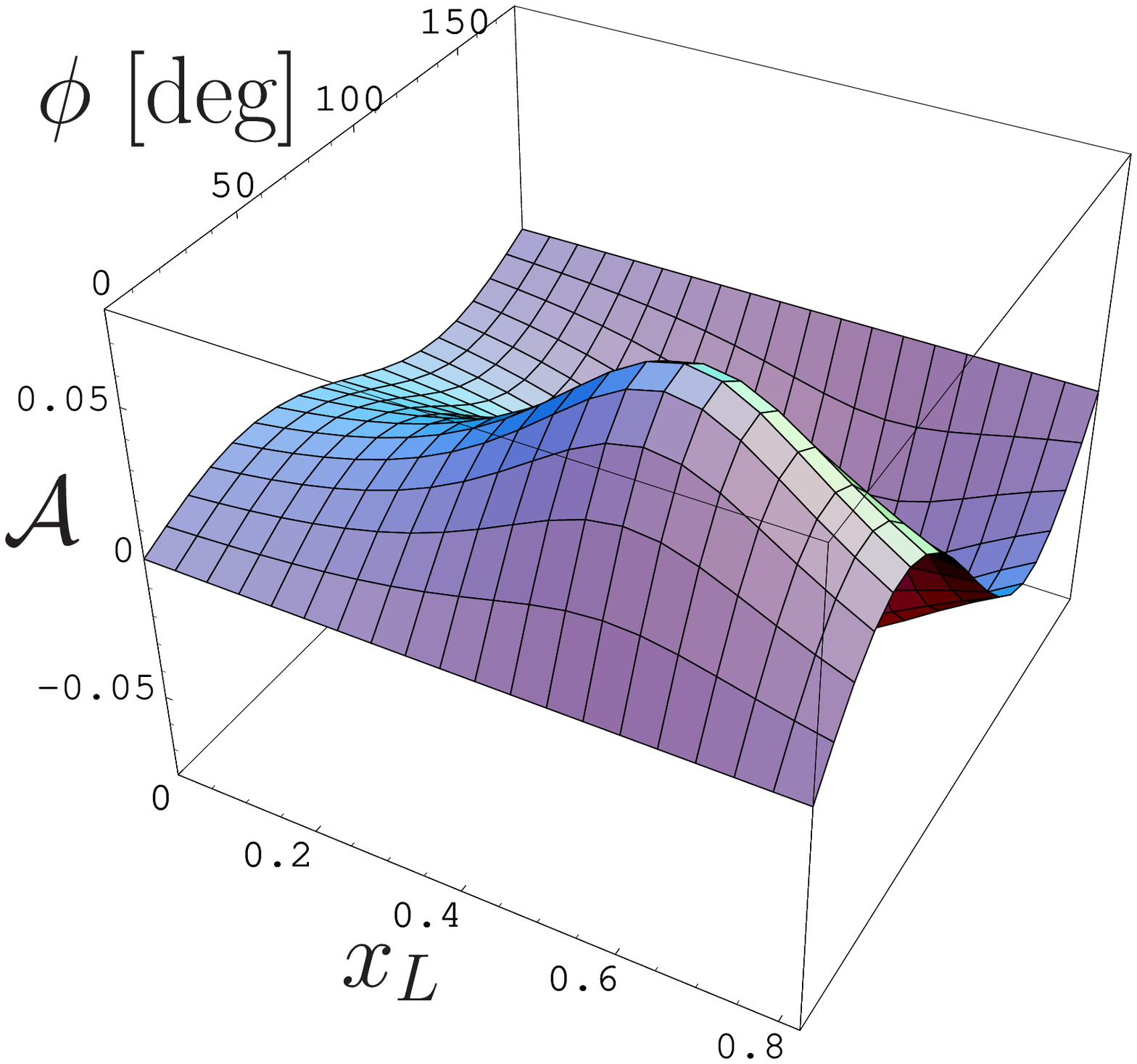}~~%
   \includegraphics[width=0.30\textwidth]{
  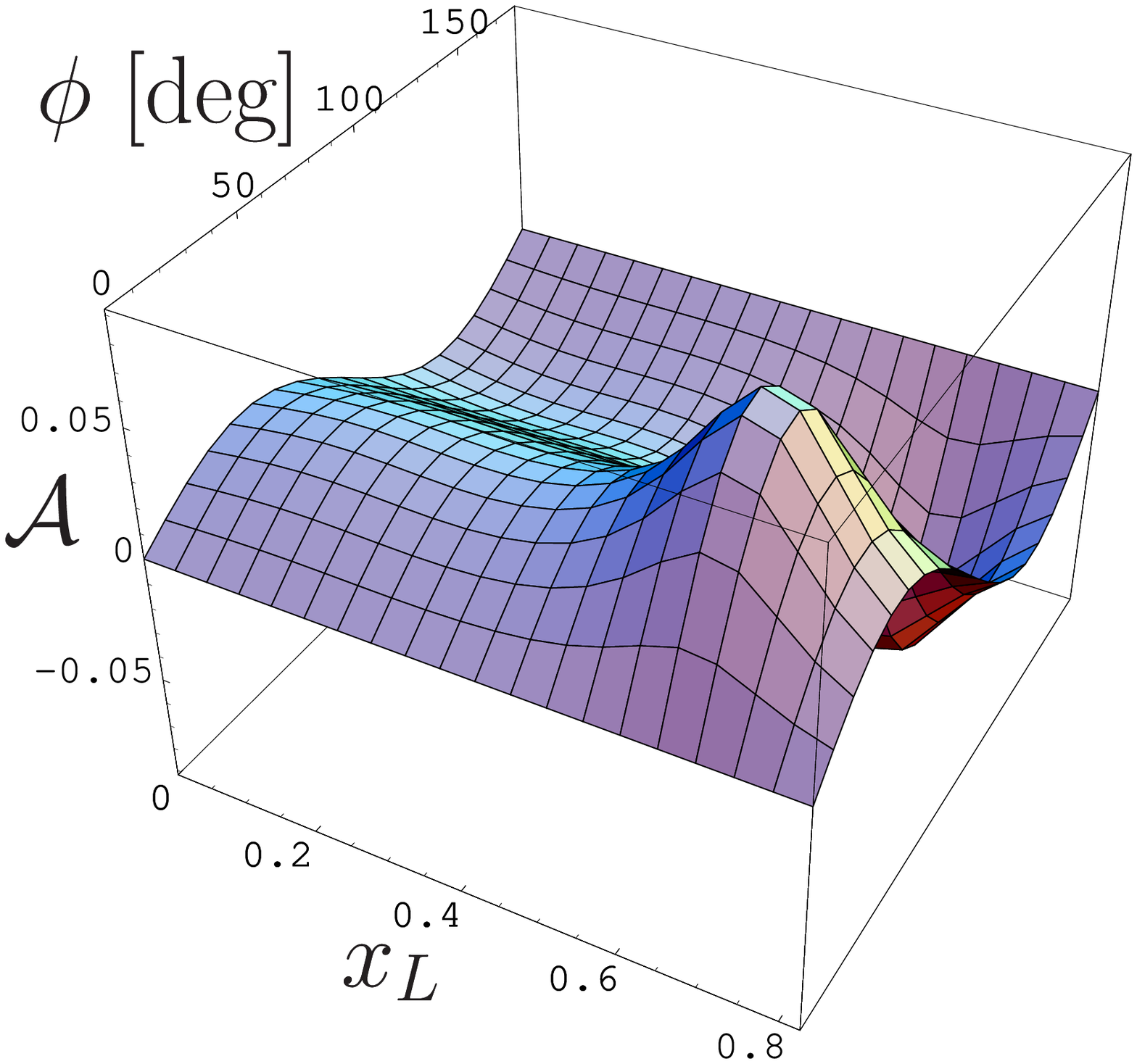}~~%
   \includegraphics[width=0.30\textwidth]{
  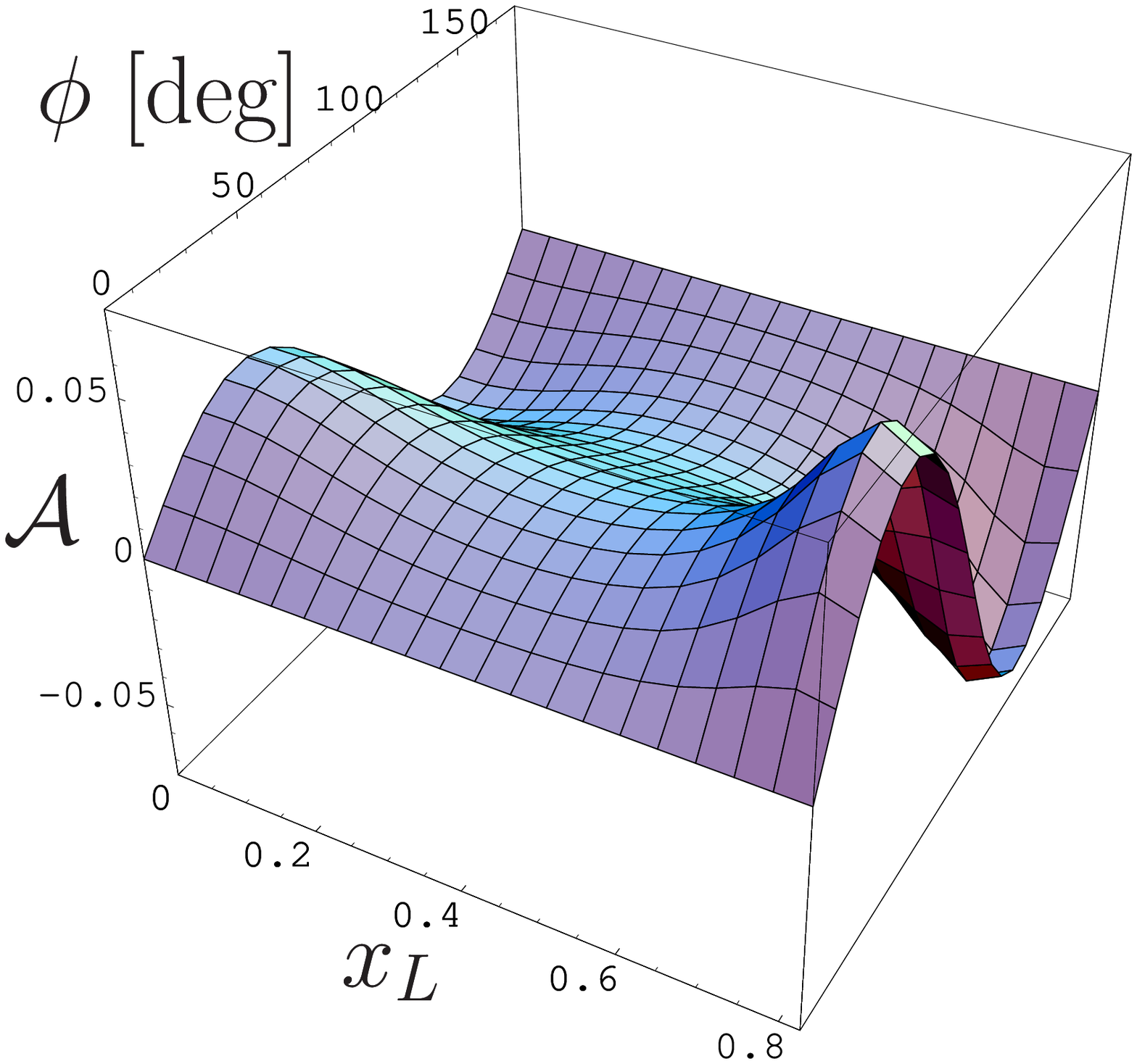}$$
 $$\includegraphics[width=0.30\textwidth]{
  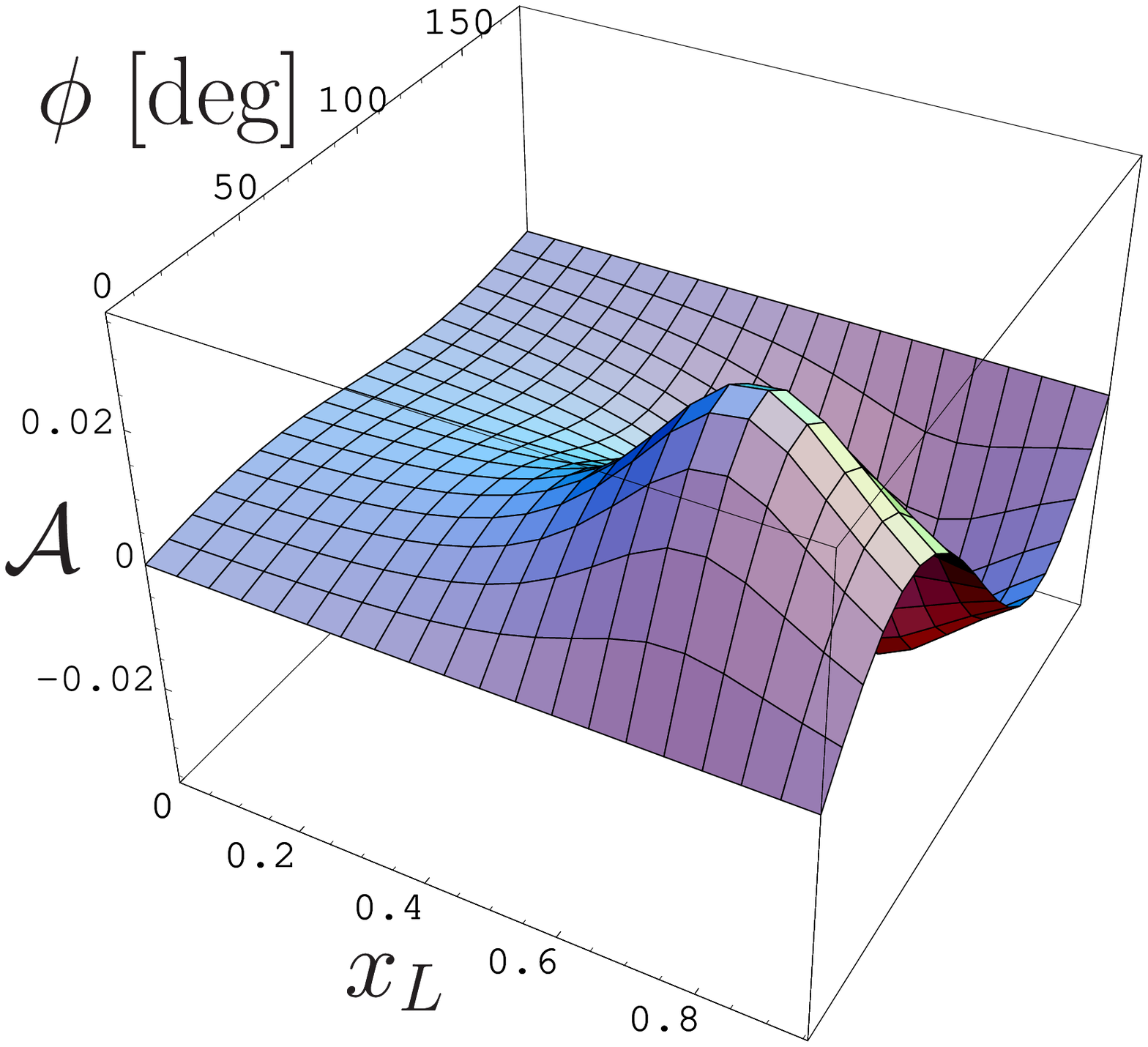}~~%
   \includegraphics[width=0.30\textwidth]{
  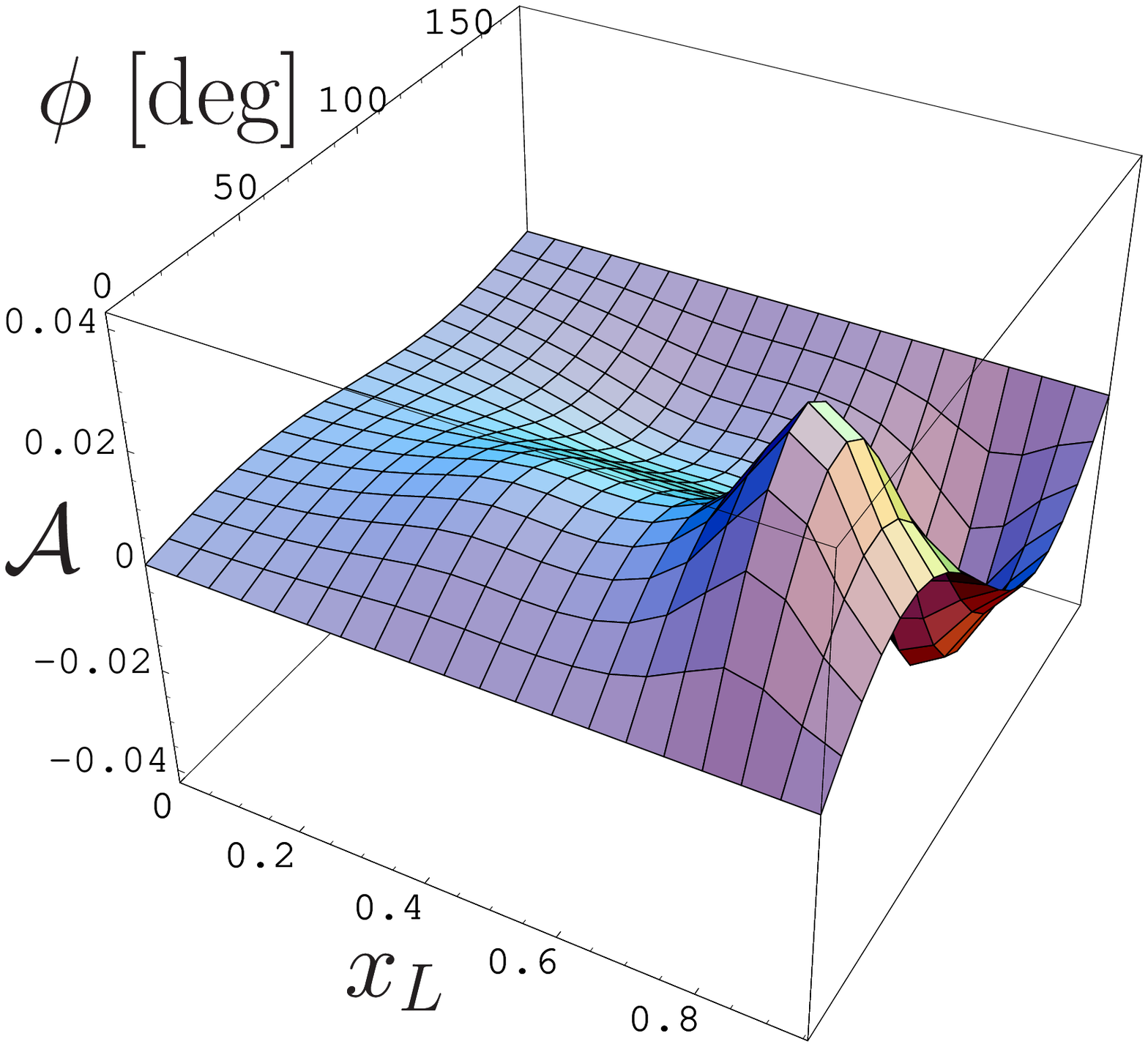}~~%
   \includegraphics[width=0.30\textwidth]{
  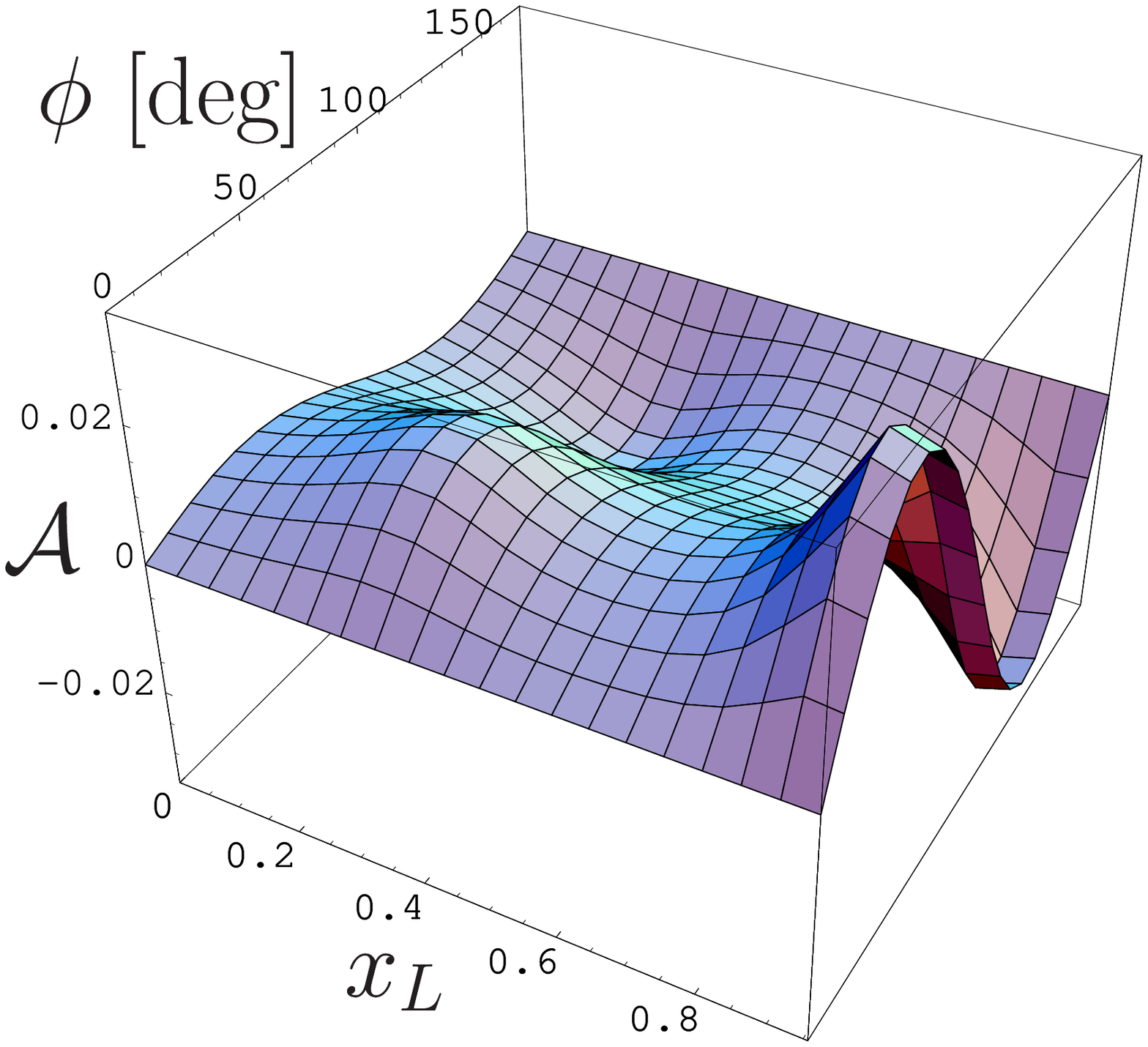}$$
   \vspace*{-7mm} \caption{3D-plots of the azimuthal asymmetry
    ${\cal A}(\phi,x_L,\rho=0.3)$. All assignments are as in Fig.\
    \ref{fig:rho-evo-2-3D}.
   \label{fig:rho-evo-Asy-030-3D}}
\end{figure}
Turn now attention to the polarized case.
Figure \ref{fig:rho-evo-1} shows the predictions for the angular
parameter $\bar\mu$ as a function of $x_L$ for three different values
of $\rho=0.06, 0.3, 0.5$ and two center-of-mass-energies
$s=100$~GeV${}^2$ (upper row) and $s=400$~GeV${}^2$ in the range
expected to be covered in the COMPASS experiment.
As before, the results shown were obtained for the asymptotic, BMS,
and CZ pion DAs.
[Similar results (not shown) hold also for the parameter $\nu$,
which is interlinked with $\mu$.]

The result of LO evolution of the pion DA as well as of the nucleon
structure functions in the angular parameter $\bar\mu$ is the detailed
image of curves shown in Fig.\ \ref{fig:rho-evo-2}.

To unravel the behavior of the various pion DAs in the endpoint region,
we also examine the angular moment (cf.\ Eq.\ (\ref{eq:Az-Mom-Def}))
and display the results in Fig.\ \ref{fig:rho-evo-Ang-Mom}.
As mentioned in the previous section, this quantity was proposed 
in \cite{BMT95} as a sensitive measure for the angular modulation 
of the pion DA at large $x_L$.
One sees from this figure that this seems indeed to be the case,
especially for not too small $\rho$ values and, once experimental data
will become available, it could potentially serve to discriminate
among different pion DAs.

To complete the analysis, we include 3D-plots of the parameter
$\bar\mu$ as a function of $x_L$ and $\rho$
(Fig.\ \ref{fig:rho-evo-2-3D}) and analogously 3D plots of the
angular asymmetry ${\cal A}(\phi, x_L, \rho )$ versus $x_l$ and $\phi$
(Fig.\ \ref{fig:rho-evo-Asy-030-3D}), the latter for $\rho=0.3$.
Note that both quantities have been evaluated for two different values
of the center-of-the-mass energy $s=100$~GeV$^2$ (upper rows) and
$s=400$~GeV$^2$ (lower rows), 
as expected for the COMPASS experiment \cite{BDP04}.

The observed increase of $\cal A$ with $x_L$ suggests that the
single-spin asymmetry of the muon-pair angular distribution is
associated with the valence Fock-state contributions in the pion DA.
This behavior is valid for all considered pion DAs and does not
significantly depend on the value of the scaling parameters $\rho$
and $\tau$, though its size decreases with increasing $s$.
We have verified that the results do not change significantly for
larger $\rho$ values.
If future experimental data would confirm such a single-spin asymmetry,
this would point to a new mechanism for generating a new nontrivial
phase in QCD factorization leading to T-odd spin asymmetries that
mimics the effect of a true T violation.

\section{Conclusions}
\label{sec:concl}
Our objective in this work was 
(i) to update previous results on the unpolarized 
    $\pi^{-}N\to\mu^{+}\mu^{-}X$ DY process 
and 
(ii) to make detailed predictions for the angular distribution 
    parameters for the hard-scattering of pions on longitudinally 
    polarized protons.
The single-spin asymmetry, predicted here for various pion 
distribution amplitudes, may soon become amenable to experimental 
check at COMPASS.
To this end, we have presented an updated analysis of the DY process 
with the inclusion of the pion's DA, the latter based on the 
theoretical appraisal of the theoretical situation obtained within 
the context of nonlocal QCD sum rules and supported by recent 
high-precision lattice calculations and other experimental data from 
the pion-photon transition.
Though the existing data on the unpolarized DY 
$\pi^{-}N\to\mu^{+}\mu^{-}$ process 
cannot single out one particular pion DA with little ambiguity, 
the ``bunch'' of the BMS DAs, derived from nonlocal QCD sum rules, 
seems to comply most favorably with the E615 data.
Given the distinctive behavior of all considered pion DAs with 
respect to the longitudinal momentum fraction, carried by the 
annihilating quark from the pion in the polarized DY process, 
one may hope that measuring the angular moment in the planned 
COMPASS experiment may lend quantitative support for one or the 
other proposed pion DA.

On the other hand, it is also important to consider another 
(and actually the most common) mechanism of the $\pi+N$ Drell--Yan 
process, treating the pion structure in terms of parton distributions
rather then the pion DA.
This mechanism is the dominant one at moderate values of $x_L$.
There are also related contributions to the SSA due to the imaginary 
phases emerging in the short-distance subprocesses~\cite{PR83,CW92}.
These contributions have recently been studied for the kinematics 
of RHIC and J-PARC~\cite{Yok07}.
The investigation of the relevance of this mechanism in the case
of the COMPASS kinematics requires further investigation.

\acknowledgments
We would like to thank A.\ V.\ Efremov and S.\ V.\ Mikhailov for
discussions and useful remarks.
We are grateful to R.\ Bertini, F.\ Bradamante, and O.\ Denisov for
discussions on the COMPASS experiment and its potential extension 
to the Drell--Yan process.
Two of us (A.P.B. and O.V.T.) are indebted to Prof.\ Klaus Goeke
for the warm hospitality at Bochum University,
where part of this work was carried out.
N.G.S is grateful for support to BLTP@JINR, 
where this work was completed.
O.V.T. is also indebted to M.~Anselmino, F.~Balestra, R.~Bertini,
A.~Kotzinyan, and G.~Pontecorvo
for the warm hospitality and useful discussions at INFN (Torino).

This work was supported in part by 
the Deutsche Forschungsgemeinschaft, grant 436 RUS 113/881/0, 
the Heisenberg--Landau Programme (grant 2007),
the Russian Foundation for Fundamental Research, 
grants No.\ 06-02-16215 and No.\ 07-02-91557,
and the Russian Federation Ministry of Education and Science 
(grant MIREA 2.2.2.2.6546).

\end{document}